\documentclass{aa}
\usepackage[varg]{txfonts}
\usepackage{graphicx}
\usepackage{subfigure}
\usepackage{longtable}
\usepackage{xcolor}
\usepackage{dblfloatfix}
\usepackage{natbib}
\bibpunct{(}{)}{;}{a}{}{,} 

\makeatletter
\let\do@linenumbers\relax
\let\linenumbers\relax

\makeatother

\def\0{\hspace*{0.5em}}
\begin{document}

\title{Rotation period of the O giant $\xi$ Persei: Coexistence of cyclic and periodic variability 
}
\subtitle{ A weakly magnetic candidate star}
\author{H.\,F.~Henrichs\inst{1}\corrauth{h.f.henrichs@uva.nl}
\and N.\,P.~Sudnik\inst{1
} \and  A.~David-Uraz\inst{
2}
}

\institute{
Anton Pannekoek Instituut voor Sterrenkunde, Universiteit van Amsterdam,
Science Park 904, 1098 XH Amsterdam, {The Netherlands}, 
\and
Department of Physics, Central Michigan University,
Mount Pleasant, MI 48859, USA
}

\date{Received 25 January 2026 / Accepted 12 June 2026}


\abstract
{OB stars 
exhibit line profile variability,  
often over timescales associated with rotation; notably, in their ultraviolet wind-sensitive lines
where `discrete absorption components' tend to recur cyclically. }
{We searched for periodicity in spectral regions of the O7.5III(n)((f)) star \object{$\xi$\,Persei} formed very close to the star and least likely to be affected by doublet overlap or irregular transient phenomena.
}
{We 
identified the low-velocity wind 
region 
probed by the \ion{N}{IV}
$\lambda$1718\,$\AA$\ line 
as the most uncontaminated spectral region for which a large and homogeneous dataset is available:  
307  International Ultraviolet Explorer (IUE) spectra  
over 12 years and 11 STIS {(on board the {\it Hubble} Space Telescope)} spectra taken 21 years later. 
We also studied 322 time-resolved \ion{He}{II} $\lambda$4686\,$\AA$\ spectra. 
We searched for periodicities in high-precision space photometry  (MOST, BRITE, and TESS), covering a time span of 13 years. We also reconsidered X-ray studies {with \it  Chandra}.
}
{A {\tt CLEAN} analysis and subsequent weighted least-squares fit of the 
flux in this region 
resulted in a unique period of $2.040514 \pm 0.000020\,$d, attributed to rotation. 
The phase of maximum 
flux in the \ion{N}{IV},  \ion{Si}{IV},  
H$\alpha$, and \ion{He}{II} line profiles coincide with the results of X-ray variability studies. 
We also found 
this period with low amplitude in the 
photometric data,  
albeit with large intrinsic scatter, and with a deviating maximum phase.  
}
{
Given the coherent periodic behaviour of several observables probing the region of the wind nearest the stellar surface, we propose that $\xi\,$Per has 
an as-yet undetected weak global surface magnetic field.  
We excluded a rotation period of $\sim$ 4\,d based on stellar parameters. The sinusoidal behaviour suggested that only one magnetic pole is visible, implying an inclination of $i\,\sim\,51^{\circ}$, and  therefore, $\beta$  {$\lesssim$} $90^{\circ} - i =  39^{\circ}$ for the  magnetic obliquity. 
We present a conceptual framework within which weak wind confinement by a global magnetic field can create a small magnetospheric disc, while allowing for the formation of more classic `corotating interaction regions' at higher magnetic latitudes. The periodic variations are then understood to result from the changing projected area of the magnetospheric disc as a function of rotational phase, both along the line of sight and off the stellar limb. 
}

\titlerunning{Rotation period of the O giant $\xi$ Persei}

\authorrunning{H.~F.\ Henrichs, N.\ Sudnik, A.\ David-Uraz}

\maketitle
\nolinenumbers

\section{Introduction}\label{sect:intro}
Many spectral lines in OB stars show unexplained variability on a rotational timescale. This occurs, for example, in the so-called discrete absorption components (DACs) in 
ultraviolet (UV) wind line profiles and in many optical lines 
(e.g. \citealt{Henrichs:1988, Prinja:1988, massa:1995, kaper:1996, kaper:1999, fullerton:2003, fullerton:2011}).
This variability is considered to be cyclical 
(e.g. in the case of sunspots), rather than periodic, meaning that it can be phased over short timescales, similarly to periodic variations, but not over long timescales.
Often, the variability pattern can be recognised over a few consecutive rotational cycles, but it usually changes when comparing datasets that are years apart.
The absence of strict periodicity is in accordance with the lack of evidence of a ubiquitous 
stable, globally organised surface magnetic fields among OB stars.  Among a sample of 
97 O-type stars, only about 7$\%$ of them were found to be magnetic oblique rotators with 
large-scale fields that generally exhibit simple geometries (dominated by a dipolar component; \citealt{grunhut:2017}). The remaining stars were found to yield non-detections with typical upper limits of the order of 300\,G in the majority of cases. These fields are understood to be fossil in origin, meaning that they are not contemporaneously generated or maintained by dynamo activity, rather being relics of an earlier event or evolutionary stage (e.g. \citealt{borra:1982}) and found to be stable over timescales of years to decades \citep{silvester:2014}. Examples of magnetic O stars include the O7Vp star \object{$\theta^1$\,Ori\,C} \citep{donati:2002}, the Of?p star \object{HD\,191612} \citep{donati:2006a}, the O9.5Ia star \object{$\zeta$\,OriAa} 
\citep{bouret:2012, blazere:2015}, which is the O-type star with the weakest polar field strength known yet, 
and most recently the O8II((f)) star \object{63\,Oph \citep{barron:2025}}.

Time-resolved UV spectra, predominantly obtained with the International Ultraviolet Explorer (IUE) satellite \citep{Boggess:1978}, which was active from 1978 to 1996, are available only for a few of these magnetic stars. In addition, finding periodicity in the wind lines of these stars is often difficult because of the lack of sufficient time coverage. In attempts to identify the cause of the wind variability and its connection to the rotation of the underlying star, knowledge of the rotation period of the star itself is imperative, but absent for the stars without a detected magnetic field. One particularly challenging example is the O7.5III(n)((f)) star $\xi\,$Persei (\object{HD\,24912}) for which excellent UV datasets obtained with IUE are available in the form of densely sampled time series, each spanning several to 10 days and spread over 12 years. These are often accompanied by a campaign with simultaneous ground-based spectroscopic observations \citep{henrichs:1994b, kaper:1997, dejong:2001}. The UV line profile variations, mainly in the \ion{Si}{iv} $\lambda\lambda 1394, 1402\,\AA$\ doublet, suggest a periodicity of around 2 or around 4 days \citep{kaper:1996, kaper:1999}. Such a period is unlikely to be associated with non-radial pulsations, which have much shorter periods, 
of the order of hours \citep{dejong:1999}. The most likely origin is therefore the rotation period, which would provide a stable clock in the case the star hosts a stable magnetic (dipolar) field, or would produce cyclic behaviour if transient surface perturbations are present (similar to transient spots). The estimated stellar radius combined with its $v \sin i$ value indicates indeed a value of the rotation period in the observed range 
with a wide margin.
The existing magnetic measurements of this star could only provide an upper limit of  $B_{\rm p} \le$ 59\,G for the polar field strength, as reported by \cite{david-uraz:2014}. It should be noted that this upper limit lies below the observed lower limit for large-scale dipolar field strengths among massive stars and, instead, it lies within the so-called magnetic desert \citep{auriere:2007}. 

Ultimately, \cite{david-uraz:2014}  concluded that there are no observational grounds for the assumption that 
stable, large-scale magnetic fields are behind the recurrent behaviour of the UV DACs in $\xi\,$Per. 
None of the 13 investigated O stars for which sufficient UV spectroscopic and optical spectropolarimetric datasets were available were detected to host such a field. The most accepted model for the origin of DACs is that their origin is nevertheless connected to localised stellar surface phenomena  \citep{cranmer:1996}, which cause co-rotating interaction regions (CIRs), as suggested by \cite{mullan:1984, mullan:1986}, which are observed as DACs. The physical cause of the perturbation at the stellar surface according to \cite{cranmer:1996} could be equally due to surface magnetic fields or non-radial pulsations, in both cases leading to similar wind structures. \cite{david-uraz:2017} applied this model to the star $\xi\,$Per, with different bright-spot properties, and could reproduce the basic structures as observed in UV spectral time series with a number of given spot models. As these authors argue, this scenario is compatible with a stochastic, as opposed to a permanent, nature of such spots, which may appear and disappear, thereby explaining the cyclical rather than periodic behaviour.  In this respect, the long-term space photometry presented by \cite{ramiaramanantsoa:2014} indicated variations in the light curve in the 2\,d period range, which the authors interpreted as being caused by several stochastic co-rotating bright spots on the stellar surface. Such spots were also proposed to be responsible for line profile variations observed in optical spectroscopy by \citet{britavskiy:2023}.

With the \textit{Chandra} X-ray Observatory, \cite{massa:2014} observed a 15$\%$ variation in the X-ray 
flux over the course of  45\,h, which could be modeled either by a linear decay or by a cosine curve for which they adopted a 2.086\,d period, taken from \cite{dejong:2001}. They found that the equivalent width (EW) in the blue wing of their contemporaneously taken H$\alpha$ absorption profiles had its maximum value  about 17\,h after the X-ray maximum when the cosine fit was used. Since the hardness of the X-ray spectrum did not change, they favoured the explanation that the variation of the X-rays was due to obscuration by intervening material. They offered two explanations for the region where the X-rays were formed: either distributed over the whole surface close to the star or at the trailing edges of the CIRs. The X-ray spectral line analysis of \textit{Chandra} spectra of $\xi\,$Per by \cite{cohen:2014} and their subsequent modelling of the H$\alpha$ line showed that the X-ray emission region starts around 1.5\,$R_{\star}$, whereas the H$\alpha$ emission (observed on top of the photospheric absorption profile) would be expected to form just above the photosphere, the same region of the clumping onset. However, their modelling accounts for a spherically symmetric wind, an assumption that might need to be challenged, as further discussed in Sect.\,\ref{section:discussion}.
 
 New contemporaneous {\it Hubble}  Space Telescope (HST) STIS spectra, containing the \ion{Si}{IV} and \ion{N}{IV} wind lines, and {\sl XMM-Newton} observations were presented by \cite{massa:2019}, which they interpreted in terms of a 2.086\,d period from \cite{dejong:2001}.  They found that the X-ray flux was anti-correlated with the softness ratio, as opposed to what was expected. A reexamination of both 
 these results, as well as results from earlier \textit{Chandra} observations \citep{massa:2014}, is presented below. 

Our goal is to critically evaluate the large body of existing UV and optical time-resolved datasets to identify spectral regions that probe the region around the star very close to the surface and to carry out a 
period analysis. We also want to compare our findings with magnetic, photometric and X-ray studies of the star aiming at a more or less consistent picture.

Section\,\ref{section:parameters} summarises the stellar parameters. In Sect.\,\ref{section:datasets}, we detail the properties of the employed datasets. Section\,\ref{section:period} describes the result of the period analysis, which we interpret in Sect.\,\ref{section:rotation}.  
Section\,\ref{Section:photometry} shows how the inferred rotational period can also be used to analyse broadband optical space photometry.
Section\,\ref{section:magnetic} summarises all magnetic measurements so far. Extended discussions are given in Sect.\,\ref{section:discussion}. A summary of our conclusions is given in Sect.\,\ref{section:conclusions}.

\section{Stellar parameters}
\label{section:parameters}

\begin{table}[t!]
\caption[Stellar parameters]{Adopted parameters for \object{$\xi\,$Per  used in this study.}}
\centering
\begin{tabular}{llc}
\hline
\hline
 Parameter& Value & Reference\\
\hline
Spectral Type                         & O7.5\,III\,(n)((f))               & 1 \\
$V$ (mag)                                   & 4.026                             & 2 \\
$d$ (pc)                              & 418$^{+58}_{-39}$                 & 3\\
$T_{\rm eff}$ (K)                     & 33500 $\pm$ 2000                  & 4 \\
$R_{\ast} (R_{\odot})$                  & 11.2                              & 5, Sect.\,\ref{section:parameters} \\
$\alpha$ (mas)                        & 
0.245 $\pm$ 0.010
& 6 \\
$v_{\rm rad}$ (km\,s$^{-1}$)          & 65.4 $\pm$ 0.8                    & 7 \\
$v \sin i$ (km\,s$^{-1}$)              & 215                               & 8 \\
$v_{\rm mac}$ (km\,s$^{-1}$)          & 80                                & 8 \\
$v_{\rm tot}$ (km\,s$^{-1}$)          & 230                               & 
8 \\
$P_{\rm rot}$ (d)                     & 2.040514 $\pm$ 0.000020             & Sect.\,\ref{section:period}\\
$i$ ($^{\circ}$)                                   & $51$                      & Sect.\,\ref{section:rotation}\\
$\dot{M}$ $(M_{\odot}\,{\rm yr}^{-1})$   & 2.2$^{+0.6}_{-0.5}\times 10^{-7}$ & 9 \\
$v_{\infty}$ (km\,s$^{-1}$)           & 2400 $\pm$ 50                     & 10 \\
$M$ $(M_{\odot})$                         & 26                                & 11 \\
\hline
\vspace{-0.6cm}
\label{tab:parameters}
\end{tabular}\\
\tablebib{(1) \citet{sota:2011}; (2) 
\cite{maiz-apellaniz:2004}; (3) {\it Gaia} DR3 \cite{Bailer-Jones:2021};
(4) \citet{martins:2017}; (5) \citet{markova:2004}; (6) angular diameter \citep{gordon:2018}; (7) \citet{gontcharov:2006};
(8) $v \sin i$ and $v_\mathrm{mac}$ taken from \citet{david-uraz:2014}, with $v_\mathrm{tot}$ computed as the quadratic sum of the two; (9) derived by \cite{cohen:2014} based on $R = 14$  $R_{\odot}$, and $v_{\infty} = 2450$ km\,s$^{-1}$. A more consistent value would be lower since $\dot{M} \propto R v_{\infty}$; (10) \cite{kaper:1996}; (11) \cite{repolust:2004}.  Note: parameters without the uncertainties are provided in the referred papers or sections.
}
\end{table}

Table \ref{tab:parameters} gives a compilation of the adopted stellar parameters. We note that the spectral type by  
\cite{martins:2017} is given as O7.5\,III\,(n)(f), which is likely an oversight. The adopted distance of  
418$^{+58}_{-39}$ pc was  derived from the  {\it Gaia} DR3 parallax, which is fairly consistent with 416$^{+116}_{-74}\,$pc by \textsc{Hipparcos}, and with $486 \pm 57\,$pc derived from the interstellar \ion{Ca}{II} absorption \citep{megier:2009}. 
We note that the actual value of the distance is not used, as the relevant parameter is the stellar radius. Various values for the radius were derived from atmospheric modelling, which include 12\,$R_{\odot}$ \citep{leitherer:1988}, 25.5\,$R_{\odot}$ \citep{puls:1996}, 14\,$R_{\odot}$ \citep{repolust:2004}, and  11.2 or 25.2\,$R_{\odot}$ by \cite{markova:2004}. We note that the larger values of the radius are a direct consequence of the questionable reclassification of luminosity class III to I as preferred by \cite{herrero:1992},  which was based on a non-LTE analysis rather than spectral morphology. 
The distance constraints from {\it Gaia} give between 0.23 and 0.28\,mas for 11.2\,$R_{\odot}$, and between 0.51 and 0.63\,mas for 25.2\,$R_{\odot}$, respectively. Recent interferometric measurements of the star showed an angular diameter of 0.220 to 0.233\,mas with an uncertainty of 0.015\,mas \citep{gordon:2018}, which implies a radius of $10.2 \pm 1.3\ R_{\odot}$ based on an adopted distance of $433 \pm 46\,$pc.  This  quoted distance is the average of three equally weighted independent distance measurements  based on \textsc{Hipparcos} data, interstellar \ion{Ca}{II} absorption and absolute magnitude. Since the individual values and relative weights are quantitatively difficult to assess, we decided to use the value of $11.2\ R_{\odot}$ derived by \cite{markova:2004}, which we further adopt in this paper to be consistent with the other stellar parameters,  including the adopted distance within its uncertainties.  We note that the uncertainty in the radius has a strong impact on the derived inclination angle.  $\xi\,$Per is a well-known runaway star \citep{Blaauw:1961, Gies:1986}, as apparent from its high radial velocity and in particular its peculiar proper motion \cite{Maiz-Apellaniz:2018}, and is surrounded by a bow shock \citep{Bodensteiner:2018}.
\cite{Turner:2008} reported a 9\,mag fainter companion at a 2.4" separation, which was however not recovered by \cite{Maiz-apellaniz:2010}. The runaway and likely single status of $\xi\,$Per combined with its high rotation rate is in line with the findings of \cite{Britavskiy:2024}.
We adopted the value of $v \sin i$ from \cite{david-uraz:2014}, who applied the method by \cite{simon-diaz:2007} to split the total observed width of the absorption lines ($v_{\rm tot}$) into the contributions of macroturbulence ($v_{\rm mac}$) and rotational broadening ($v \sin i$), which are 
summed quadratically, see also the discussion by \cite{simon-diaz:2014}.
The rotation period is derived in Sect.\,\ref{section:period} and argued in Sect.\,\ref{section:rotation}, which includes the derivation of the inclination angle.

\section{Observational data}
\label{section:datasets}

Notorious spectral signatures that originate close to the stellar surface in the lower wind region are found in the spectral lines of \ion{Si}{IV} $\lambda$1402\,\AA\  
and \ion{N}{IV} $\lambda$1718\,\AA\ (covered by 326 IUE satellite spectra over 12 years), as well as H$\alpha$, and \ion{He}{II} $\lambda$4686\,\AA\ (covered during several campaigns, see for instance \citealt{dejong:1999, dejong:2001}). 
In fact, $\xi\,$Per has the largest number of IUE spectra among all O stars. \cite{massa:2019} presented 11 spectra containing the \ion{Si}{IV} $\lambda$1402\,\AA\  and \ion{N}{IV} $\lambda$1718\,\AA\  lines obtained with STIS on board the HST \citep{Kimble:1998}. 
We consider these 
datasets in turn.

\subsection{UV spectra}
\subsubsection{IUE}

\paragraph{(i) INES versus\ MAST wavelength scale}

\begin{figure}[!t]
\begin{center}
\includegraphics[width=0.65\linewidth]{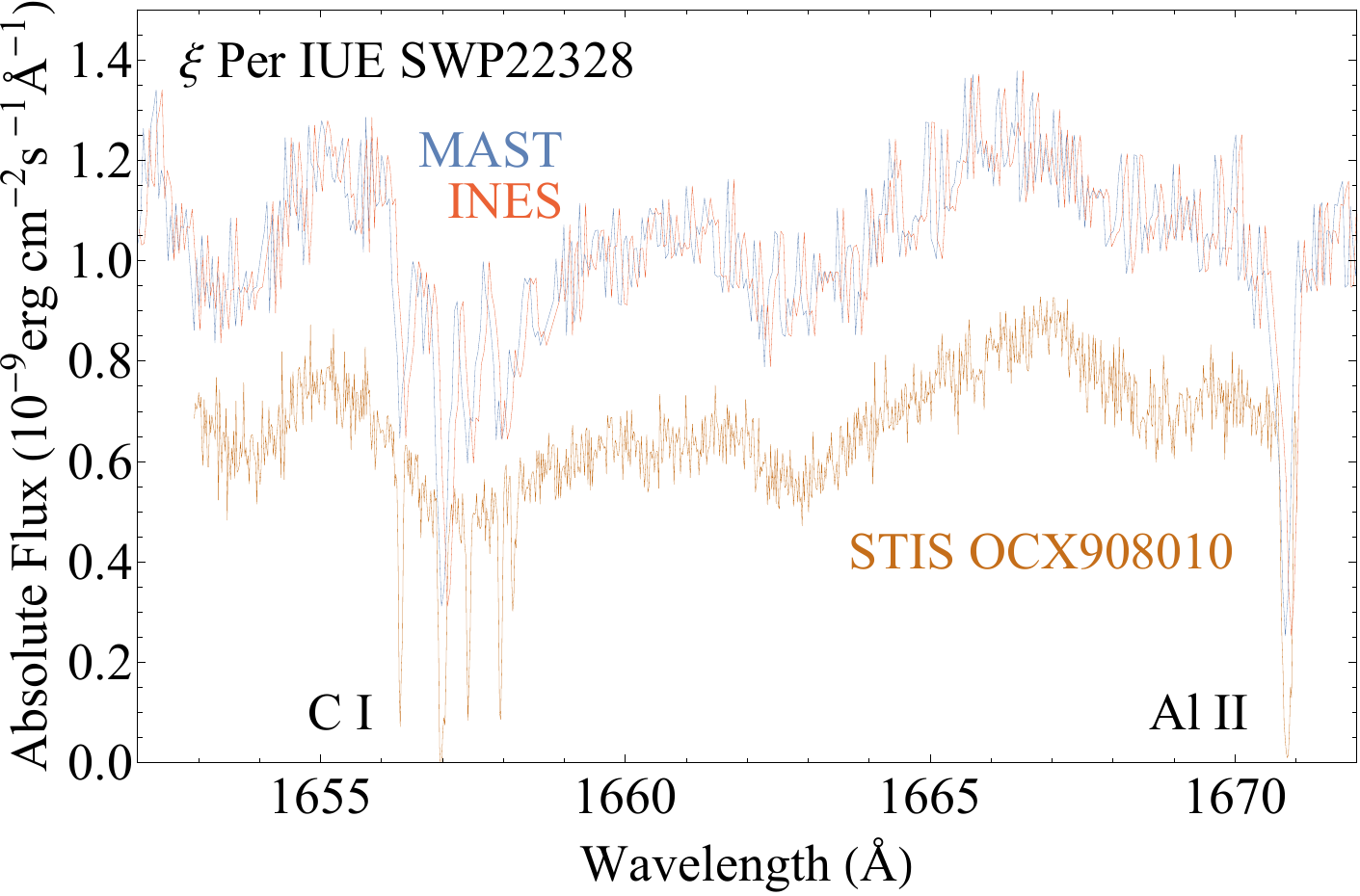}
\caption{Comparison  between spectra of $\xi\,$Per in the region around the interstellar \ion{C}{i} $\lambda$1657\,\AA\ and \ion{Al}{ii} $\lambda$1671\,\AA\ lines: STIS {(brown), IUE MAST (blue) and IUE INES (red)} of an arbitrarily chosen SWP spectrum. The STIS and MAST wavelength scales match closely, as opposed to the INES wavelength scale, which confirms the conclusion reached by \cite{Smith:2001}. Note: the absolute fluxes of this  STIS spectrum (like all others) are lower, which is addressed in Sect.\,\ref{section:STIS}. }
\label{fig:xiPer_STIS_IUE_MAST_INES}
\end{center}
\end{figure}

\begin{figure*}[!ht]
\begin{center}
\includegraphics[width=0.7\linewidth]
{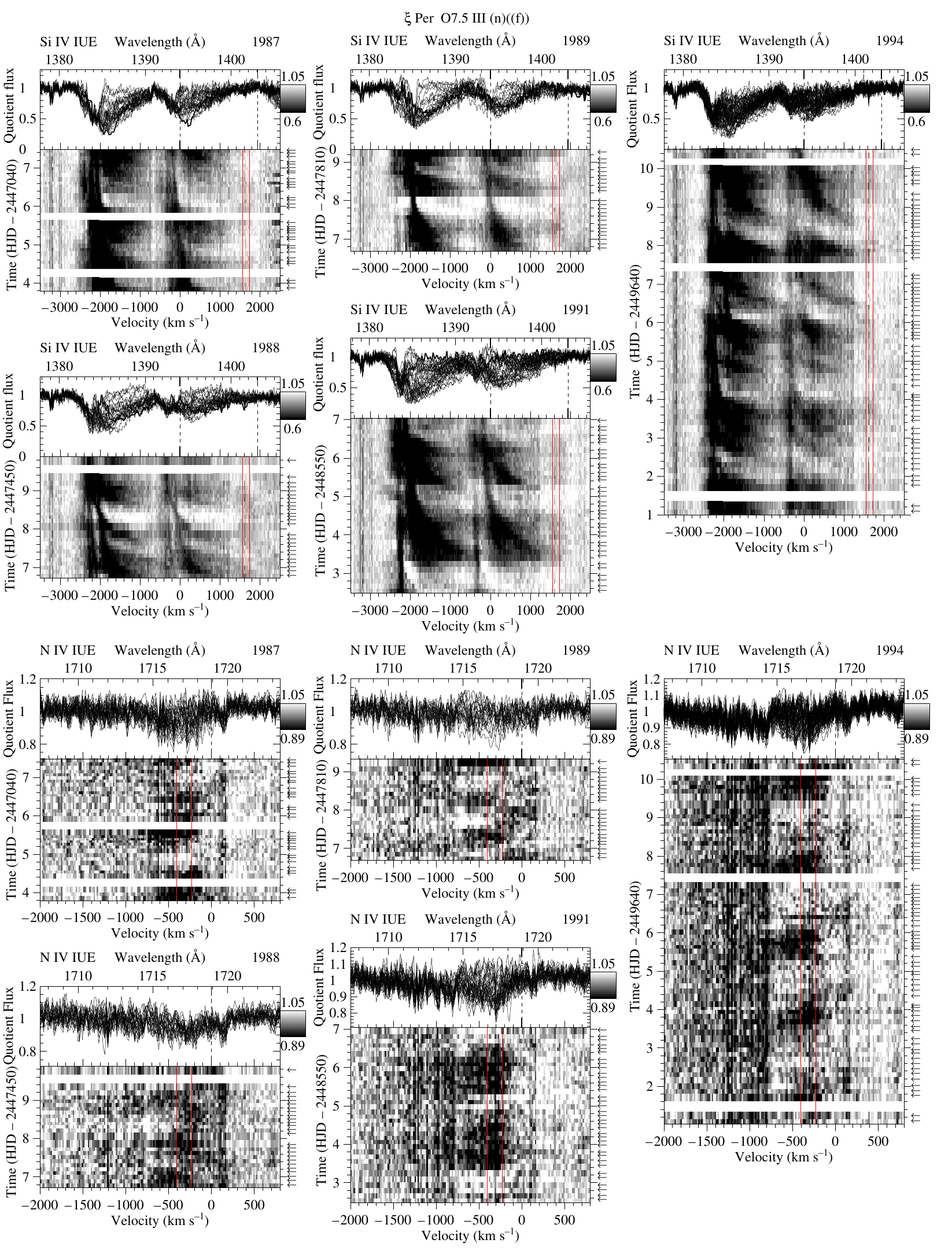}
\caption{
\ion{Si}{IV} $\lambda\lambda$ 1394, 1402\,\AA\ (\textit{top five panels}) and \ion{N}{IV} $\lambda$1718\,\AA\ (\textit{bottom five panels}) quotient line profiles, compiled from five different IUE runs obtained in 1987 (top left), 1988 (bottom left), 1989 (top middle), 1991 (bottom middle), and 1994 (right). These are constructed by dividing each spectrum by a `least-absorption' template 
\citep{kaper:1997}, and then overplotted in each top panel, while a dynamic quotient spectrum tracing the evolution of spectral features over time is shown in the corresponding bottom panel  with the epochs indicated by arrows. The \ion{Si}{IV} region best traces large-scale wind structures extending to high-velocity, demonstrating that the DACs are irregularly spaced and appear to be cyclic in nature.
The lower five panels show the same spectra in the \ion{N}{IV} region, exhibiting a qualitatively different behaviour, as the pattern in the dynamic quotient spectra appears much more repeatable. The overplotted vertical 
red lines indicate the region of interest, used for the period analysis.
}
\label{fig:xiSi4N4_87-94}
\end{center}
\end{figure*}

There are two different IUE archives, which for the high-resolution spectra taken with the  Short Wavelength Prime (SWP) camera appear to differ in wavelength scale, but with identical fluxes. All ESA IUE Newly Extracted Spectra (INES)\footnote{\url{https://sdc.cab.inta-csic.es/ines/}} SWP spectra are shifted by +17.7\,km\,s$^{-1}$ with respect to the spectra in the MAST archive. According to \cite{Gonzalez-Riestra:2000}, this shift was needed to match the wavelengths in the overlapping region of the short- and long-wavelength cameras, but \citet{Smith:2001} could not confirm this.  A comparison of the positions of the interstellar \ion{C}{i} $\lambda$1656\,\AA\ and \ion{Al}{ii} $\lambda$1671\,\AA\ lines in the IUE and STIS spectra shows that the MAST wavelengths match better than the INES wavelengths, see Fig.\,\ref{fig:xiPer_STIS_IUE_MAST_INES}. 
As we are only interested in comparing STIS and IUE spectra, we use the MAST wavelength scale with the INES 
error bars and flags for unreliable data points. 

\paragraph{(ii) Variability of wind-sensitive spectral lines}

\begin{figure*}[ht!]
  \centering
    \subfigure[]{\label{fig:IUENIV}\includegraphics[width=0.32\textwidth]{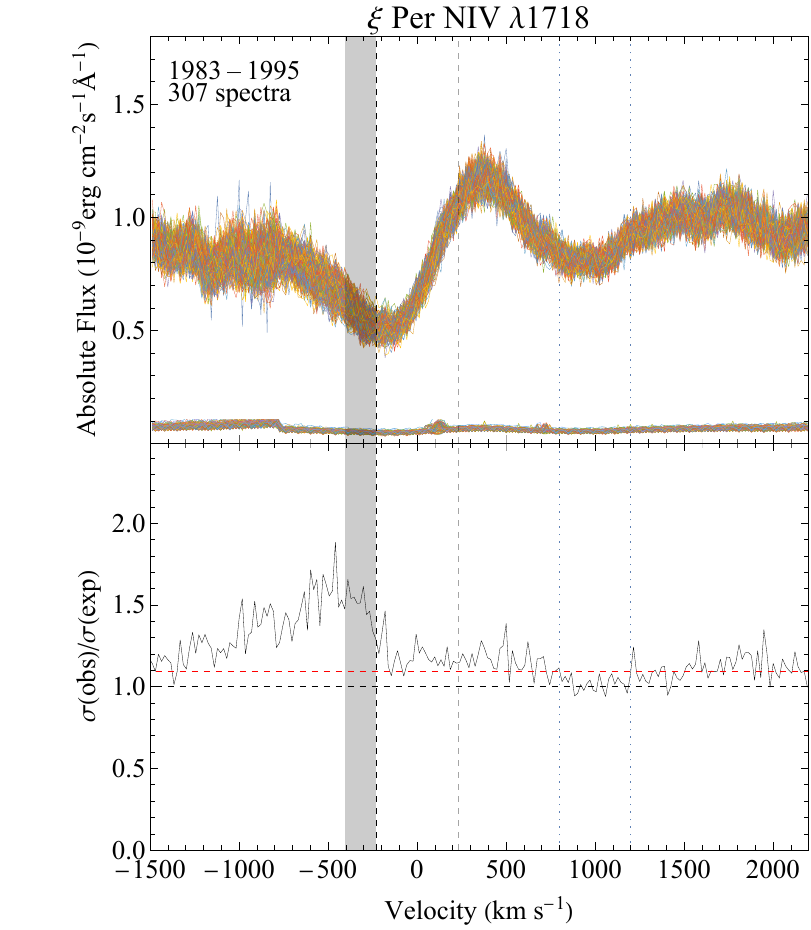}}
    \subfigure[]{\label{fig:IUESiIV}\includegraphics[width=0.32\textwidth]{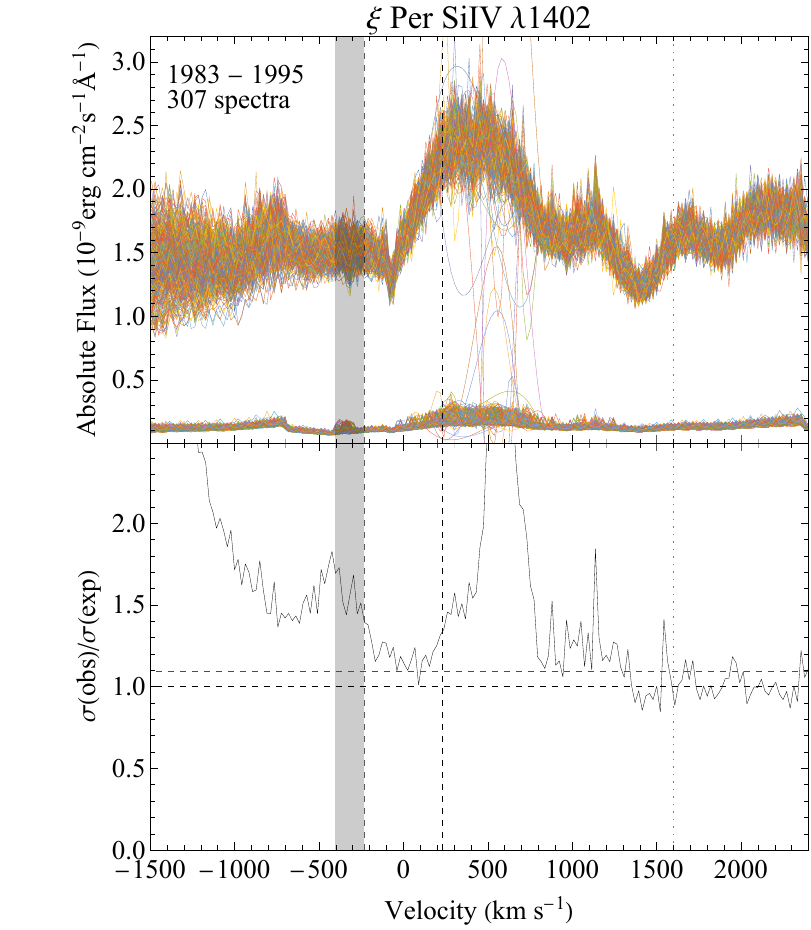}}
    \subfigure[]{\label{fig:IUENIVSIV}\includegraphics[width=0.32\textwidth]{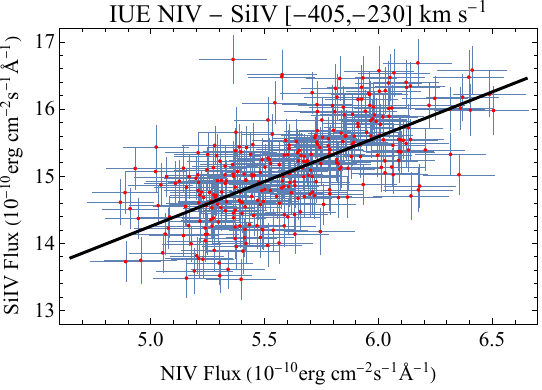}}
\caption{
\textit{Panel (a), top}: Overplot of 307 IUE spectra (listed in Table \ref{tab:IUEtable}) around \ion{N}{IV} $\lambda$1718\,\AA, taken during 12 years. The corresponding errors are plotted near the bottom. The discontinuity near $-800\,$km\,s$^{-1}$ is due to the transition to the adjacent order. The vertical dashed lines are drawn at $\pm v_{\rm tot}$ = 230\,km\,s$^{-1}$. The vertical dotted blue lines indicate the region used for the scaling. The shaded area denotes the velocity range [$-$405, $-$230]\,km\,s$^{-1}$ (the region of interest) within which the mean flux was measured.  \textit{Panel (a), bottom}: TVS as a measure of the variability.  The dashed red line indicates the 99$\%$ confidence level.
\textit{Panel (b}): Same as (a) but for  \ion{Si}{IV} $\lambda$1402\,\AA. The erratic fluxes and excess in the TVS around 650\,km\,s$^{-1}$ are due to overexposure in 33 spectra, and are outside the ROI.
\textit{Panel (c)}: Measured fluxes in the region of interest of panels\,(a) and (b), 
displaying a strong correlation.
A linear fit is overplotted. Any correlated intrinsic variation is expected to take place along the straight line.
}
\label{fig:IUEtvs}
\end{figure*}

The DAC progressions are most prominent in the \ion{Si}{IV} $\lambda\lambda$1394, 1402\,\AA\ doublet \citep{kaper:1996}, with the full dataset shown in Fig.\,\ref{fig:xiSi4N4_87-94} (the top five panels) 
including dynamic spectra for each IUE run. Since the terminal velocity of the wind is larger than the doublet separation, the low-velocity 
interval 
[$-405, -230$]\,km\,s$^{-1}$ (our region of interest) of the (strongest) blue component at $\lambda$1394\,\AA\ is contaminated with high-velocity material of the red component, making the blue component not suitable for our goal, and hence we will consider only the $\lambda$1402.770\,\AA\ component.
The singlet \ion{N}{IV} $\lambda$1718.551\,\AA\ line 
shows clear variation in concert with the DACs, which is most clearly shown in Fig.\,\ref{fig:xiSi4N4_87-94} (the lower five subpanels).

For our analysis, we used only the spectra taken in the large aperture, as spectra taken in the small aperture
are known to suffer from calibration issues.
The list of the epochs of the observations is conveniently given in Table \ref{tab:IUEtable}.
Figure\,\ref{fig:IUEtvs} shows an overplot of the remaining 307 spectra around the \ion{N}{IV} $\lambda$1718\,\AA\ line in velocity scale with respect to the stellar rest frame, along with the temporal variance spectrum (TVS;
\citealt{fullerton:1996}), used in the form of the ratio of the observed to the expected standard deviation \citep{henrichs:1994a}.  The latter was determined in the region between 800 and 1200\,km\,s$^{-1}$  in which there is no variability expected and in which the average signal-to-noise ratio {(S/N)} is around 23. The TVS ratio equals unity when no significant variability is present.

IUE spectra are known to suffer from imperfect echelle-order overlap corrections, 
specifically
between $-1300$ and $-700$\,km\,s$^{-1}$, visible as increased flux errors in Fig.\,\ref{fig:IUEtvs} 
(top left). 
The vertical dashed lines are drawn at $\pm v_{\rm tot}$. The variability above the 99$\%$ confidence 
level extends from $\simeq -1300$ to $-100$\,km\,s$^{-1}$, of which the affected region should be excluded, leaving the region $\simeq -700$ to $-100$\,km\,s$^{-1}$ as significantly variable. In Fig.\,15 in \cite{dejong:2001} the phase dependency is presented as a function of radial velocity, which shows that the phase in the \ion{N}{IV}, \ion{Si}{IV} and H$\alpha$ remains constant in the interval [$-405
$, $-230$]\,km\,s$^{-1}$. To minimise the possible contribution to the noise in phase, we therefore limited our period search to the velocity bins in the interval [$-405$\,km\,s$^{-1},-v_{\rm tot}$], i.e. the region which probes the wind layer to about 0.12\,$R_\ast$ above the surface for a $\beta = 0.8$ velocity law. Further justification for the choice of this interval is given in Sect.\, \ref{section:period}.
We measured for each spectrum $n$ the mean flux $F_ n$ within this interval, each flux point $i$ weighted by $1/\sigma_{i}^2$, with $\sigma_{i}$ its corresponding error bar. The values of $F_{\rm n}$ and $\sigma_{F_ n}$  are listed in Table\,\ref{tab:IUEtable}.

The result of a similar approach for the \ion{Si}{IV} $\lambda$1402\,\AA\ line is presented in Fig.\,\ref{fig:IUESiIV}.  The applied scaling region is [1600, 2200]\,km\,s$^{-1}$.
Figure\,\ref{fig:IUENIVSIV} shows that the measured \ion{N}{iv} and \ion{Si}{iv} fluxes are strongly correlated, which confirms the physical relation between the line-forming regions in this velocity interval.

\begin{table}[!ht]
\caption[]{{Journal of the 307 IUE observations, available in electronic form at the CDS (Sect.\,\ref{electronic}).}}
\centering
\begin{tabular}{l@{\enspace}l@{\enspace}l@{\enspace}l@{\enspace}l@{\enspace}l@{\enspace}l}
\hline
\hline
SWP & Date           & HJD        & Flux        & $\sigma_{\rm Flux}$ & Flux         & $\sigma_{\rm Flux}$ \\
    &     & $-$2440000 & \ion{N}{IV} & \ion{N}{IV}         & \ion{Si}{IV} & \ion{Si}{IV} \\
\hline
18941   &       1983-01-04      &       5338.64218      &       5.29    &       0.19    &       1.52    &       0.03    \\
20606   &       1983-08-04      &       5551.21402      &       5.72    &       0.18    &       1.56    &       0.03    \\
20607   &       1983-08004      &       5551.23436      &       6.21    &       0.17    &       1.53    &       0.03\\
\hline
\label{tab:IUEtable}
\end{tabular}\\
\vspace{-0.5cm}
\tablefoot{First three columns: image number taken with the Short Wavelength Prime (SWP) Camera,  date of the observation in the format (yyyy-mm-dd), and heliocentric Julian date (HJD)  at the centre of the exposure, respectively. Last four columns: flux and its 1-$\sigma$ error bar in units of $10^{-10}$ erg\,s$^{-1}$cm$^{-2}$\AA$^{-1}$ for \ion{N}{iv} $\lambda$1718\,\AA\ and in units of $10^{-9}$ erg\,s$^{-1}$cm$^{-2}$\AA$^{-1}$ for \ion{Si}{IV }$\lambda$1402\,\AA, respectively, measured in the interval [$-405,-230$] \,km\,s$^{-1}$.} \\
\end{table}

\begin{figure*}[ht!]
  \centering
    \subfigure[]{\label{fig:STISNIV}\includegraphics[width=0.32\textwidth]{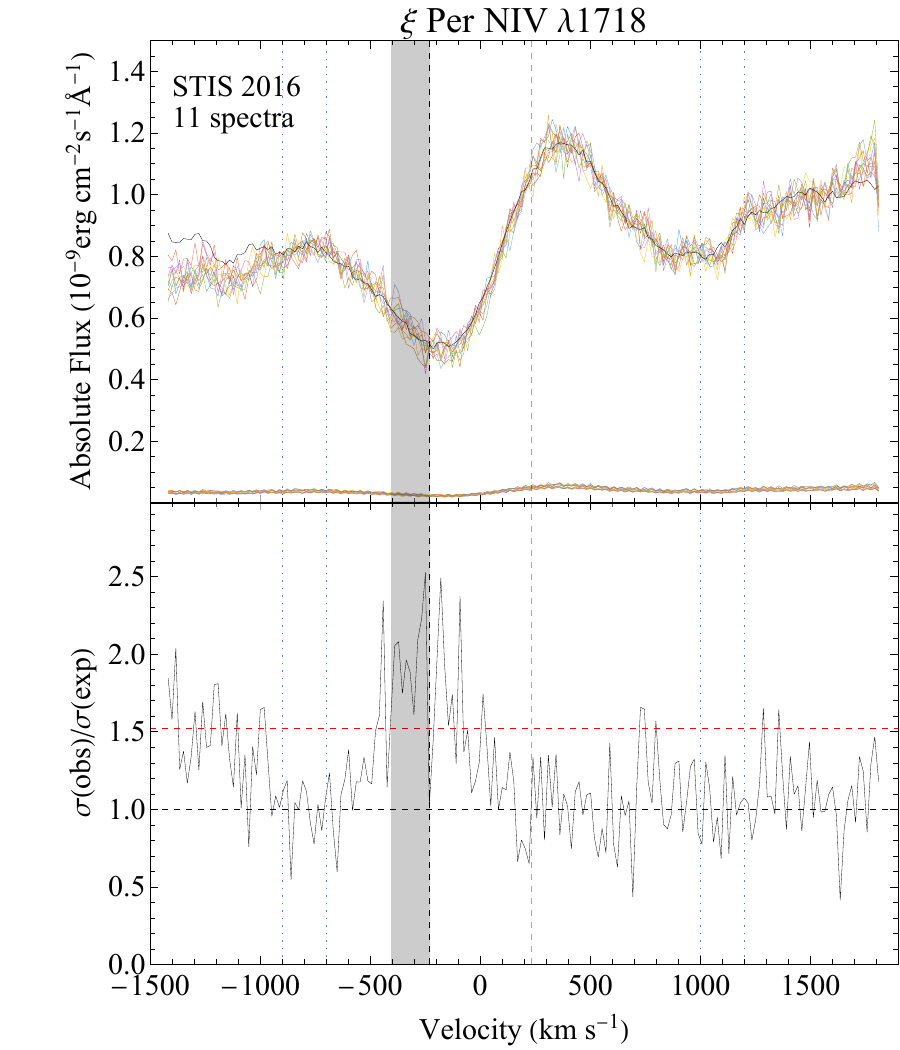}}
    \subfigure[]{\label{fig:STISSiIV}\includegraphics[width=0.32\textwidth]{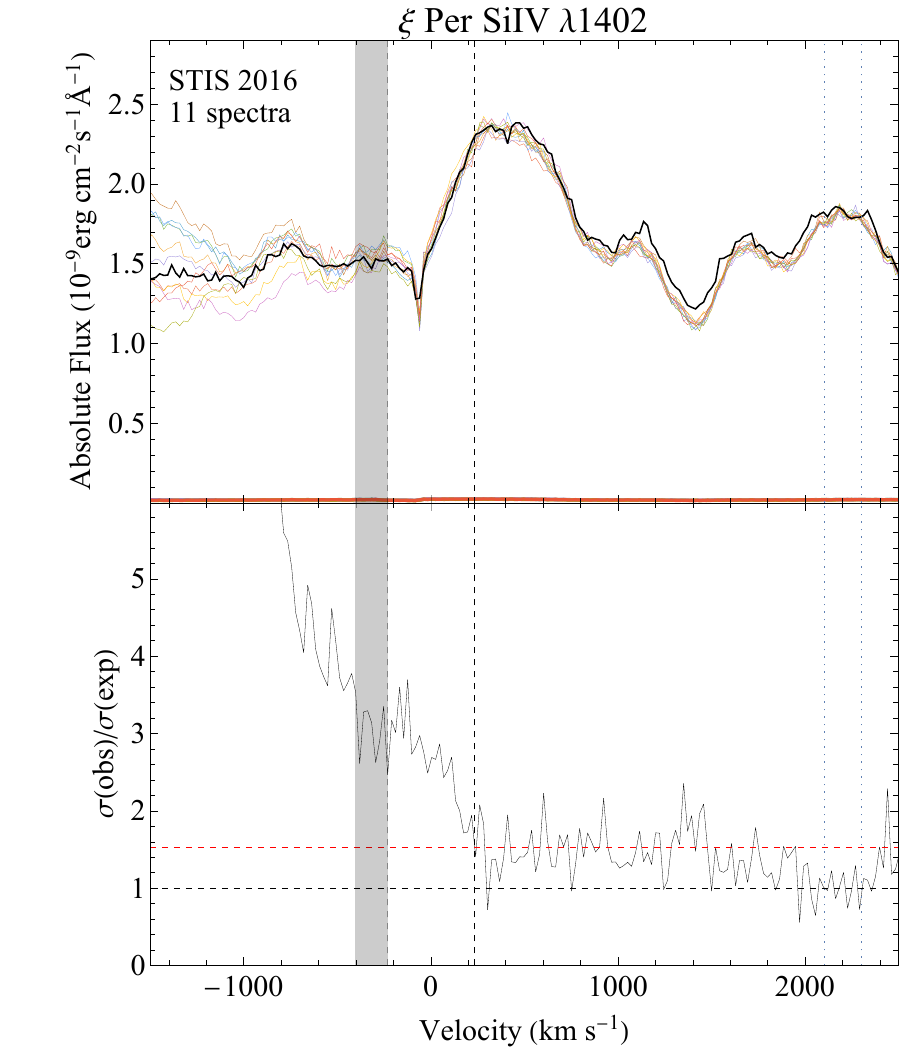}}
    \subfigure[]{\label{fig:STISNIVSIV}\includegraphics[width=0.32\textwidth]{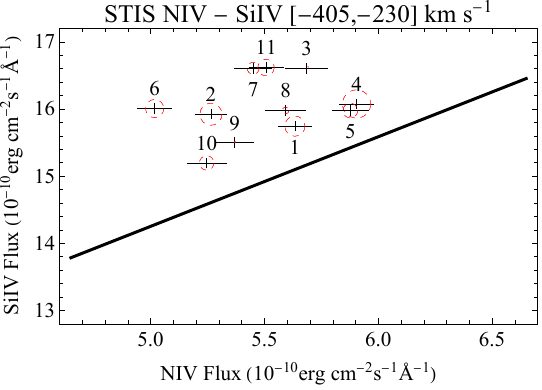}}
\caption{
Same as Fig.\,\ref{fig:IUEtvs} but for the 11 STIS spectra. (a) \ion{N}{IV} region. 
The mean
of the IUE spectra is overplotted as a thick black curve, indicating a good match. 
(b) Similar for the \ion{Si}{IV} region, indicating a significant mismatch. (c) Mean fluxes with formal error bars over the same ROI of the \ion{Si}{iv} against \ion{N}{iv} lines for the 11 STIS spectra, plotted with the same axes as in Fig.\,\ref{fig:IUEtvs}. 
The linear fit of Fig.\,\ref{fig:IUEtvs} is overplotted. There appears to be no correlation between the two quantities, unlike the IUE fluxes.
The annotated numbers indicate the order of the spectra in time. The diameters of the dashed red circles are proportional to the S/N (Table\,\ref{tab:STIStable}), which are used as weights: for instance, numbers 3 and 9 have a very low S/N. Note that the S/Ns are not correlated to the vertical error bars, which is indicative of an instrumental problem.  Note that the \ion{N}{IV} fluxes span the same range as 
the IUE fluxes, but that the \ion{Si}{iv} STIS fluxes are systematically higher compared to IUE, 
showing that they cannot be trusted.
}
\label{fig:STIStvs}
\end{figure*}
 
\subsubsection{STIS}
\label{section:STIS}
We analysed the 11 available archival HST STIS spectra containing the \ion{N}{IV} and \ion{Si}{IV} line profiles,  obtained within four days in February 2016, 21 years after the last IUE spectrum, contemporaneously with {\it XMM-Newton} spectral and flux observations \citep{massa:2019}.  In this paper, relative fluxes were measured in the velocity region [$-$500, 0]\,km\,s$^{-1}$ in both lines and 
phased with the 2.086\,d period taken from \cite{dejong:2001}. As we derive a different period and consider a different velocity region, we reanalysed the STIS spectra. 
As emphasised by \cite{massa:2019} the mean fluxes of the 11 STIS spectra varied considerably and are 25--40$\%$ smaller as compared to the IUE fluxes (Fig.\,\ref{fig:xiPer_STIS_IUE_MAST_INES}), which is attributed to a combination of instrumental effects, such as focus, precise pointing and thermal variations. 
For confirmation, we compared IUE and STIS spectra of the only two other O stars acquired with the same EM140 grating: $\zeta$\,Pup and \object{HD\,108} which were taken in the 0.3$\times$0.05ND and 0.2$\times$0.2ND neutral density apertures, respectively, whereas $\xi$\,Per was taken in the 0.2$\times$0.05ND aperture, and found at most a few percent flux difference for $\zeta$\,Pup and HD\,108, as expected. This confirms that the IUE fluxes of $\xi$\,Per are correct, and that the effects mentioned above in combination with the used filter dominated the variation in the flux levels for $\xi$\,Per, which is not the case for $\zeta$\,Pup and HD\,108. 
Therefore, we used the average IUE absolute fluxes to match the STIS fluxes.  
Significant corrections were needed to bring them on mutually the same absolute flux scale. As a measure of the quality we take the value of the FITS key SNRMEAN (Table\,\ref{tab:STIStable}), which represents the mean value of {S/N} of good pixels. The range in values is large: for instance, the ratio between spectra 4 and 3 is nearly 60, and appears 
to be strongly related to the width of the echelle orders: some are very broad, resulting in a lower resolution as well. This difference in quality means that relative changes need to be critically examined before evaluation. 

The procedures to extract the \ion{N}{IV} and \ion{Si}{IV} are very different because the \ion{N}{IV} region is located near the edge on the MAMA Far-UV CCD in the last order without overlap with the adjacent order, and is excluded in the standard pipeline product. We describe the procedures in turn.

\begin{table*}[!ht]
\caption[]{{Journal of the 11 STIS observations, available in electronic form at the CDS (Sect.\,\ref{electronic}).}}
\centering
\begin{tabular}{l@{\enspace}l@{\enspace}l@{\enspace}l@{\enspace}l@{\enspace}l@{\enspace}l@{\enspace}l@{\enspace}l}
\hline
\hline
STIS ID   & Start time     & Exp.\,time  & HJD        &  Mean S/N & Flux   & $\sigma_{\rm Flux}$ & Flux        & $\sigma_{\rm Flux}$\\
          &     (UT)       &  (s)          & ($-$2450000) &           & \ion{N}{IV} & \ion{N}{IV}         & \ion{Si}{IV} & \ion{Si}{IV}    \\
\hline
ocx901010       &       2016-02-15      02:10:23        &       1250    &       7433.59778      &       1.95    
&       5.63    &       0.07    &       1.57    &       0.01    \\
ocx902010       &       2016-02-15      19:39:38        &       1350    &       7434.32701      &       2.12    
&       5.26    &       0.07    &       1.59    &       0.01    \\
ocx903010       &       2016-02-16      02:01:11        &       1250    &       7434.5914       &       0.046
&       5.69    &       0.09    &       1.66    &       0.01    \\
\label{tab:STIStable}
\end{tabular}\\
\vspace{-0.5cm}
\tablefoot{First four columns: STIS image identification,  observation start time in the format (yyyy-mm-dd hh:mm:ss), exposure time, and  heliocentric Julian date (HJD) at the centre of the exposure, respectively. Fifth column:  
Rounded values of the mean S/N 
of good pixels, obtained from the FITS key {\tt SNRMEAN} in the header.  Last four columns: Same as in Table \ref{tab:IUEtable}.}
\end{table*}

\paragraph{(i) \ion{N}{IV} analysis}

For the extraction of the \ion{N}{IV} line profile, we followed a nearly similar analysis by hand as described by \cite{massa:2019} and started from the photometrically and geometrically corrected data image ($*\_$flt.fits files) containing 1024 $\times$ 1024 pixels, of which 1022 columns contain useful data.
We selected a region of 23 rows which contained the last order, and determined the background in each of the 1022 columns by calculating a linear fit over the fluxes on both sides of the maximum flux. We 
excluded the region around it by first fitting a gaussian function through the flux, 
while excluding 3 times the fitted width. This was necessary since the order does not run parallel to the edge of the CCD. After subtraction of the background, the total flux was calculated by summing the pixel values.
The starting wavelength at 1710.75\,\AA\ was taken from \cite{massa:2019} with dispersion coefficient 0.0182307 \AA/pixel, which differs from the erroneous value given in the paper (D.\,Massa, priv.\ comm.).  

To enable a proper comparison with IUE spectra, a conversion to absolute fluxes is necessary. For this purpose, the (unknown) blaze function is needed, which we derived empirically in two different ways. The first method was by extracting the first neighbouring order from the *flt.fits files in the same way as for the \ion{N}{IV} order and by dividing the collected counts by the blaze-corrected 1d order, following \cite{baer:2018}. For every spectrum, this thus-obtained STIS blaze function was used to construct the \ion{N}{IV} order with absolute fluxes. However, a reasonable match of these fluxes to the same absolute flux scale of the IUE spectra could only be obtained
in a very arbitrary way,  
which led us to reject this conversion. 

The second method 
was to divide the extracted neighbouring order by the mean of the IUE spectra in the same wavelength region, 
which brings the 11 flux levels automatically to the same level. 
We applied a final scaling by selecting two small segments (indicated in  Fig.\,\ref{fig:STISNIV}), adding a 6$\%$ background, and applying a small shift (15 -- 60 pix) of this empirical blaze function such that the mean IUE and STIS spectra overlapped over nearly the whole STIS order 
(Fig.\,\ref{fig:STISNIV}). This is the calibration we adopted for further analysis for the \ion{N}{IV} line. The TVS in the lower panel of Fig.\,\ref{fig:STISNIV}  shows comparable values in the region of interest
(ROI; greyed area) with respect to the TVS of the IUE spectra in Fig.\,\ref{fig:IUEtvs}, which gives sufficient confidence in the analysis described above.  The measured fluxes are listed in Table\,\ref{tab:STIStable}.

\paragraph{(ii) \ion{Si}{IV} analysis}

The \ion{Si}{IV} doublet profiles extend over three spectral orders of the EM140 grating. Since 
our ROI 
coincides with the overlapping part of the echelle orders 20 and 19 which showed considerable flux differences, we first applied the echelle-order blaze function correction \citep{baer:2018}, before averaging the fluxes in the overlapping regions. This is not part of the archival data products, and was not applied by \cite{massa:2019}. This procedure ensures an optimised match of the fluxes in the overlapping region. 
 
To match the IUE fluxes, the mean flux
in the intervals $[-7500, -7100]$ and $[2100, 2300]$\,km\,s$^{-1}$ was selected for normalisation as they fall outside the echelle-order overlap regions of both IUE and STIS spectra, are free of suspected flux variability, and are at the highest flux levels. 
We therefore expected that within this range the fluxes would reasonably match outside the wind-affected part, but this 
was not 
the case: 
for $1000 < v < 2100$\,km\,s$^{-1}$ the STIS fluxes are significantly lower than the averaged IUE fluxes 
(Fig.\,\ref{fig:STISSiIV}),  
perhaps due to different background corrections. 
We were not able to find a better approach, for instance following the same reduction from scratch as we did for the \ion{N}{IV} order. Hence, we concluded that the standard pipeline 
could not be improved. This, however, immediately casts serious doubts about the absolute values of the fluxes in the ROI of the
\ion{Si}{IV} region, with an uncertainty that ultimately precludes a quantitative estimate, even though 
the TVS in the lower panel of Fig.\,\ref{fig:STISSiIV}  shows significant variability over a similar velocity range as for the IUE spectra in Fig.\,\ref{fig:IUESiIV}. The measured fluxes are listed in Table\,\ref{tab:STIStable}.

Figure\,\ref{fig:STISNIVSIV} shows that the measured \ion{N}{iv} and \ion{Si}{iv} fluxes of the STIS spectra are 
not correlated, as opposed to the IUE fluxes (Fig.\,\ref{fig:IUENIVSIV}). 
Since the range of the \ion{N}{IV} STIS and IUE fluxes and profiles match (Fig.\,\ref{fig:STISNIV}) we do trust the \ion{N}{IV} STIS fluxes for a combined period analysis, in contrast to the STIS \ion{Si}{IV} fluxes, as shown above. 
Since in addition, 
the quality of the \ion{Si}{IV} STIS spectra varies significantly, it does not make sense to include them 
in a period analysis.

\subsection{Optical spectroscopy: \ion{He}{II} $\lambda$4686\,\AA\ }

\begin{figure}[!t]
\begin{center}
\includegraphics[width=0.7\linewidth]{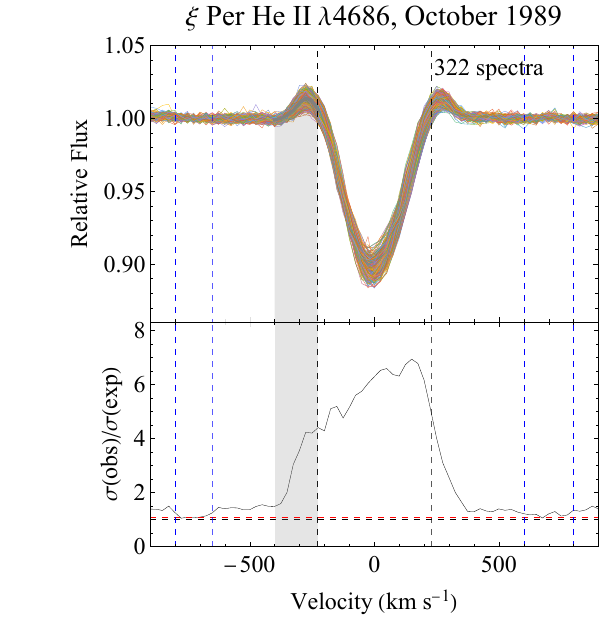}
\caption{
Overplot of 322 spectra around \ion{He}{II} $\lambda$4686\,\AA, acquired during five nights in October 1989. Similar format as in Fig.\,\ref{fig:IUENIV}. The vertical dashed blue lines correspond to the selected regions of the continuum used for normalisation. 
}
\label{fig:he2}
\end{center}
\end{figure}

\begin{figure*}[!hb]
  \centering
    \subfigure[b]{\label{fig:period04}\includegraphics[width=0.3\textwidth]  {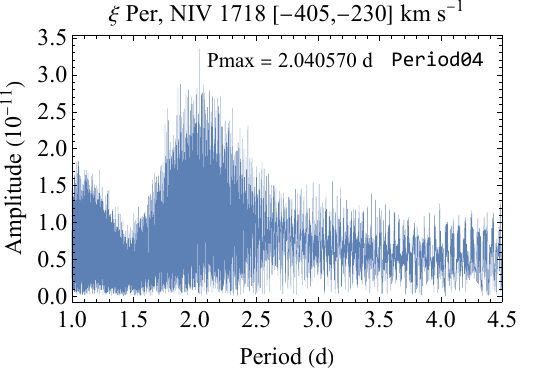}}
    \subfigure[b]{\label{fig:clean}\includegraphics[width=0.3\textwidth]{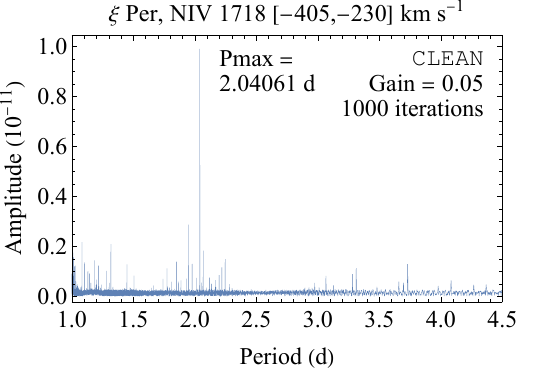}}
    \subfigure[b]{\label{fig:LombScargleIUE}\includegraphics[width=0.3\textwidth]{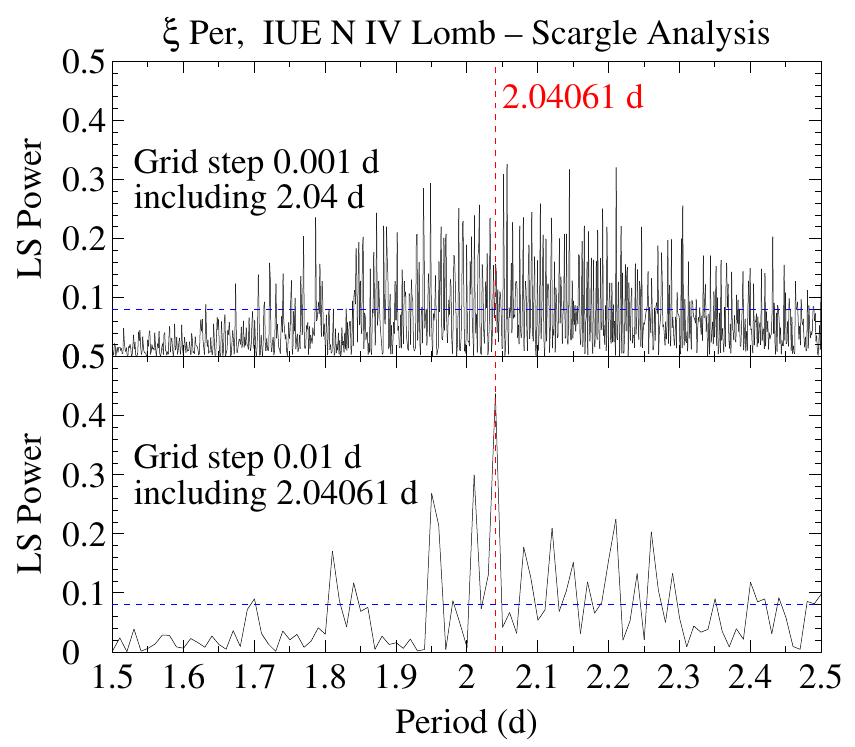}}
\caption{
Three different period-analysis results of the 307 IUE measurements from Fig.\,\ref{fig:IUEtvs}a. 
\textit{Panel (a)}: Amplitude as a function of period,  using a {\tt Period04} weighted Fourier analysis. The maximum occurs at $P_{\rm max}$  = 2.04057\,d with  {S/N =} 5.0.
\textit{Panel (b)}: Same but from a {\tt CLEAN} analysis, with maximum at 2.04061\,d.
\textit{Panel (c)}: Comparison of  generalised Lomb-Scargle periodograms showing the power based on two different frequency grids with different steps. In the top subpanel, the power calculated on a fine grid shows a multitude of non-significant peaks, despite 
exceeding the formal 99$\%$ confidence level (dashed blue line). {The GLS method failed to reveal} power at the 2.04061\,d period found by {\tt CLEAN} (dashed red line). The bottom subpanel shows the same 
power, based on a much coarser grid, but including the period found by {\tt CLEAN}. The pronounced peak exceeds the 99$\%$ confidence level by more than a factor of  5. This {contrast} underlines the critically important choice of frequency grid for the GLS  analysis in case of extremely unequally sampled data, such as in this case. 
}
\label{fig:period}
\end{figure*}

We used an extensive dataset with 322 
spectra covering the \ion{He}{II} $\lambda$4686\,\AA\ line with a {S/N} between 500 and 3000 after re-binning on a grid with 25\,km\,s$^{-1}$ bin width. This was obtained in October 1989 during 5.5 consecutive nights with a time resolution of several minutes at the Calar Alto and Kitt Peak observatories,  
collected as part of a campaign with simultaneous coverage by the IUE satellite \citep{henrichs:1994a}. The spectra used here were previously
analysed for the \ion{He}{I} $\lambda$4713\,\AA\ absorption line to search for non-radial pulsations, as described in detail by \cite{dejong:1999}.  
The \ion{He}{ii} line shows weak double-peaked emission on top of a variable absorption profile 
(Fig.\,\ref{fig:he2}). The TVS shows significant variability over a velocity range [$-$400, 400]\,km\,s$^{-1}$. Since this line also probes the inner wind just above the surface 
we measured the relative flux variations in the same velocity interval. The flux behaviour and its correlation with the UV \ion{N}{IV} line are presented and analysed in Sect.~\ref{section:period}.

\section{Period analysis}\label{section:period}

\subsection{UV data}
\label{section:priodUV}
We first  analysed the IUE data.  We used several algorithms to search for periodicity in the flux of the \ion{N}{IV} line. We found that the following methods 
gave consistent results: (1) the date compensated discrete Fourier transform algorithm by {\tt VStar}\footnote{\url{https://www.aavso.org/vstar}}, giving 2.04054\,d; (2) the discrete fourier analysis (using weights) by {\tt Period04} \citep{lenz:2004, lenz:2005, lenz:2014}, giving 2.04057\,d (Fig.\,\ref{fig:period04}); and (3) the {\tt CLEAN} algorithm by \cite{roberts:1987}, giving 2.04061\,d after 1000 iterations with gain 0.05 (Fig.\,\ref{fig:clean}). This amplitude diagram 
gave a very pronounced peak due to the nature of the {\tt CLEAN} algorithm, which removes amplitude of neighbouring smaller peaks by pre-whitening the signal with these neighbouring frequencies in favour of the predominant peak.
\begin{figure*}[ht!]
  \centering
    \subfigure[]{\label{fig:STISgroups}\includegraphics[width=0.3\textwidth]{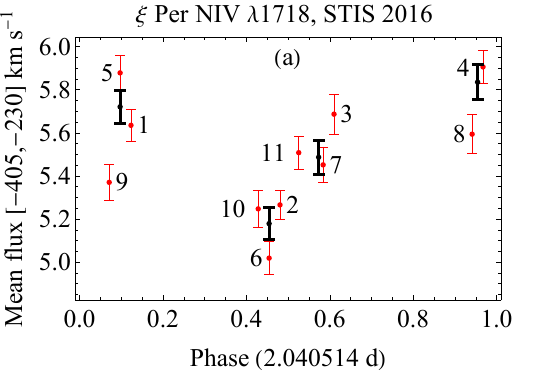}}
    \subfigure[]{\label{fig:NIVfit}\includegraphics[width=0.32\textwidth]{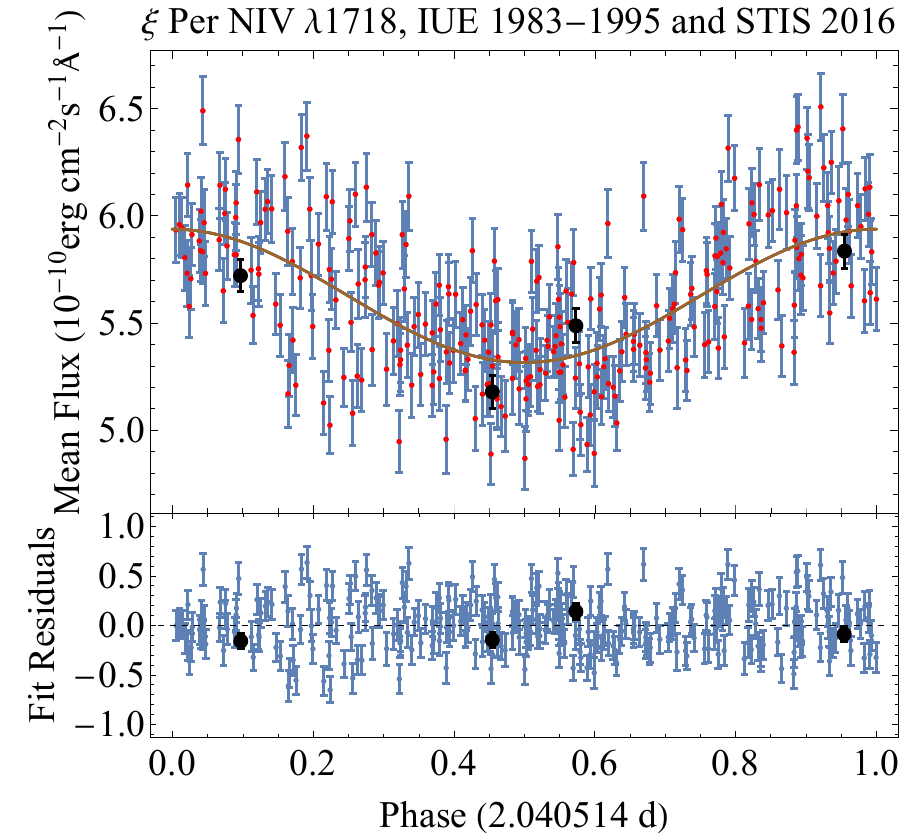}}
    \subfigure[]{\label{fig:SiIVfit}\includegraphics[width=0.32\textwidth]{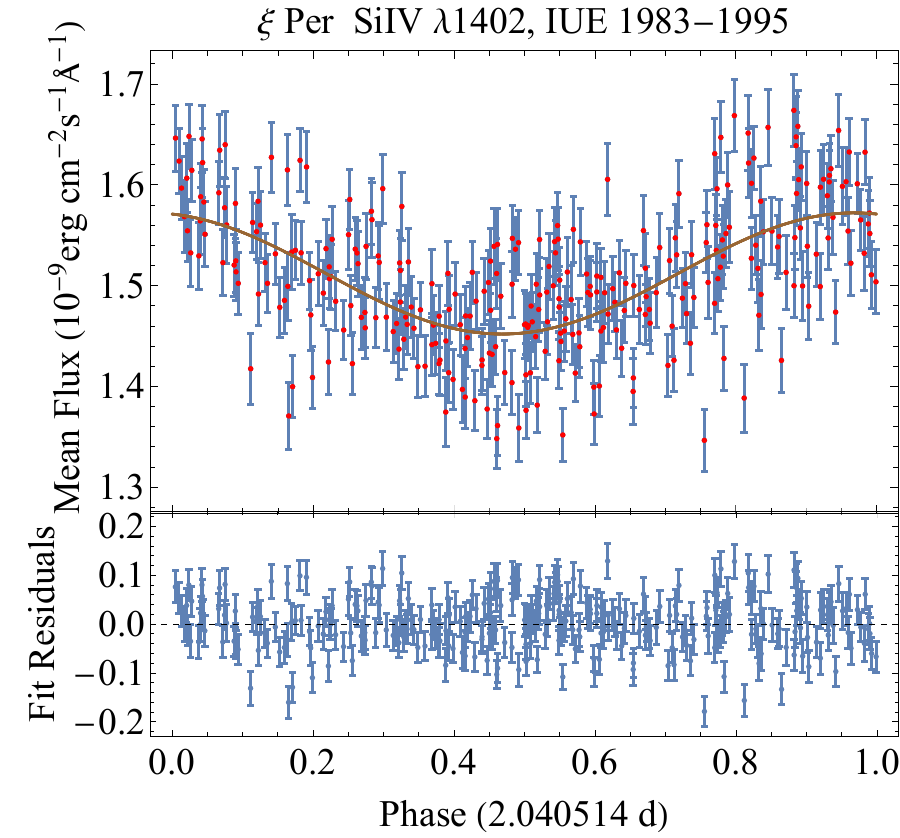}}
\caption{
\textit{Panel (a)}: Red dots show the mean fluxes in the ROI of \ion{N}{IV} STIS spectra in units of $10^{-10}$\,erg cm$^{-2}$s$^{-1}$\AA$^{-1}$ as a function of the adopted phase. Black dots show the S/N-weighted grouped points which are close in phase, which are used in the fit shown in panel\,(b), and named '4w' in Table\,\ref{tab:fitresults}. 
The numbering of the red points corresponds to the chronological order (Table\,\ref{tab:STIStable}). \textit{Panel (b), top:} Phase-folded IUE and STIS flux measurements of the \ion{N}{IV} $\lambda$1718\,\AA\ line in the ROI. 
Red dots with blue error bars: IUE data. Black points: weighted STIS data. The best weighted least-squares cosine fit is overplotted. \textit{Panel (b) bottom}: Fit residuals in the same units as in the top panel. \textit{Panel
(c)}: Same as in (b) but for \ion{Si}{IV} $\lambda$1402\,\AA\ line, IUE only.
}
\label{fig:NIVSiIVfit}
\end{figure*}

\begin{table*}[hb!]
\caption[]{Least-squares fit results for different datasets.} 
\centering
\begin{tabular}{l@{\enspace}c@{\enspace}cc@{\enspace}cc@{\enspace}cc@{\enspace}cc}
\hline
\hline
Dataset & $P$ & $\sigma_{\rm P}$& $t_0$ & $\sigma_{\rm t_0}$  & $a$ & $\sigma_{\rm a}$ & $b$ & $\sigma_{\rm b}$ &$\chi^2_{\rm red}$\\
        & (d)   &   (d)         & (HJD) & (d) &  &  &  & \\    
\hline
\ion{N}{IV} IUE (1983--1995) 307 & 2.040541 &  0.000034 & 2447565.421 & 0.023 & 0.32 & 0.02 &  5.63 & 0.02 & 3.02\\
\ion{N}{IV} IUE  307 + STIS (2016) 11 &  2.040473 &  0.000014 & 2451389.270 & 0.027 & 0.31 & 0.02 &  5.62 & 0.01 & 3.17\\
\ion{N}{IV} IUE 307 + STIS 4w  &  2.040501 &  0.000023 & 2451387.280 & 0.044 & 0.31 & 0.02 &  5.63 & 0.01 & 3.04\\
\ion{Si}{IV} IUE 307            &  2.040546 &  0.000037 & 2447565.346 & 0.025 & 0.60 & 0.04 & 15.10 & 0.03 & 2.89\\
Adopted values    &   2.040514  &   0.000020  & 2451387.302 &  0.022  & \dots & \dots &  \dots & \dots & \dots \\
\hline
\ion{N}{IV} IUE (1983--1995) 307 & 2.040514f & \dots & 2447565.425 & 0.022 & 0.32 & 0.02 &  5.63 & 0.01 & 3.03\\
\ion{Si}{IV} IUE 307            &  2.040514f & \dots & 2447565.357 & 0.024 & 0.60 & 0.04 & 15.12 & 0.03 & 2.89\\

\ion{He}{II} (1989) 322        & 2.040514f & \dots & 2447820.321 &  0.010  &  0.0040 & 0.0002  &  1.0071    & 0.00010 & 32.35  \\
H$\alpha$ (1994) 354     EW       & 2.040514f & \dots & 2449645.84  &   0.05  & \dots & \dots & \dots & \dots & \dots \\
H$\alpha$ (2004)  8    EW\tablefootmark{a}       & 2.040514f & \dots & 2453088.212 &  0.052  & 0.111   & 0.044   & 0.143   &0.016 & 1.30 \\
\textit{Chandra} (2004)   22 \tablefootmark{b}     & 2.040514f & \dots & 2453086.948 & 0.066&   35 & 6 & 525    & 5 & 0.84  \\
BRITE {(2014--2015)}   2995 \tablefootmark{c}     & 2.0389 &   0.0013        & 2456982.839 & 0.027&   0.00182 & 0.00015 & 0.00014    & 0.00011 & 11.1  \\
BRITE {(2014--2015)}   2995 \tablefootmark{c}     & 2.040514f & \dots & 2456982.842 & 0.027&   0.00182 & 0.00015 & 0.00014    & 0.00011 & 11.1  \\
\hline
\vspace{-0.6cm}
\label{tab:fitresults}
\end{tabular}\\
\tablefoot{First column: dataset with number of data points. The designation '4w' signifies a weighted mean of points in the four STIS groups with similar phase.  The fit equation is given in Eq.~\ref{eq:Cosinefit} with the same symbols as in the top line of the table. Columns 2 and 3 list the best fit period and its formal 1-$\sigma$ uncertainty.  The suffix 'f' denotes a fixed value of the period in the fit. Columns 4 and 5 list the HJD and its formal uncertainty of phase 0 of the fit. Columns 6 -- 9 list the amplitude $a$ and mean value $b$ of the fit with their formal uncertainties in units of $10^{-10}$\,erg cm$^{-2}$s$^{-1}$\AA$^{-1}$, or different units as indicated in the footnotes. The last column gives the reduced $\chi^2$ of the fit. The fifth line of the table gives the adopted values of the period and zero phase with their formal 1-$\sigma$ uncertainties, used throughout this paper, as explained in the text preceding Eq.\,\ref{eq:maxphaseNIV}.  Entries below the horizontal line were obtained with the adopted value of the period.\\
\tablefoottext{a}{Amplitude $a$ and mean value $b$ of the cosine fit with their formal uncertainties in units of \AA.}\\
\tablefoottext{b}{Amplitude $a$ and mean value $b$ of the cosine fit with their formal uncertainties in units of counts s$^{-1}$.}\\
\tablefoottext{c}{Amplitude $a$ and mean value $b$ of the cosine fit with their formal uncertainties in units of relative flux.}
}
\end{table*}

\begin{figure*}[!ht]
\begin{center}
\includegraphics[width=0.75\linewidth]
{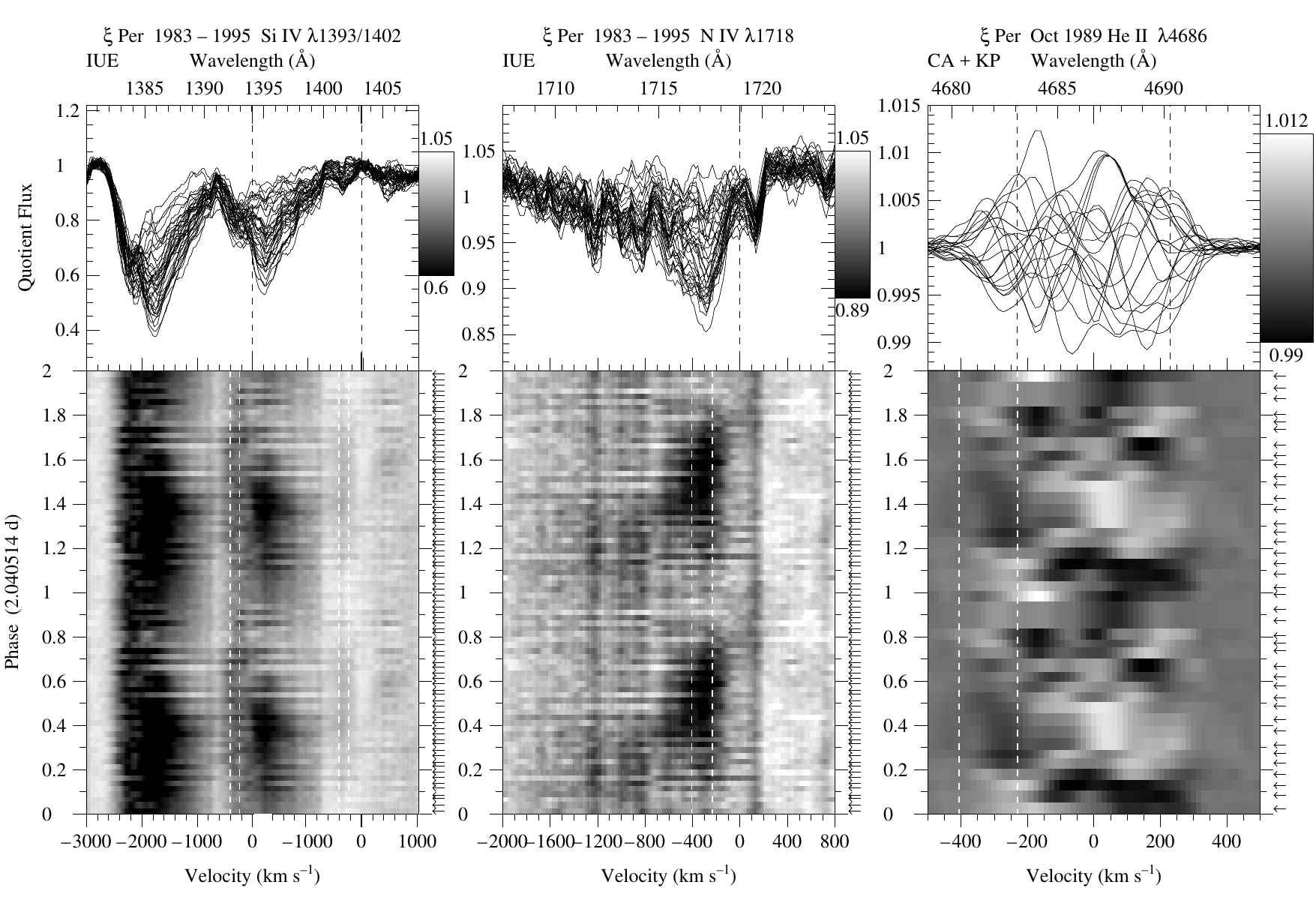}
\caption{Phase-folded dynamic plots  
of IUE spectra in the \ion{Si}{IV} and \ion{N}{iv} region during 1983-1995,  along with 322  
normalised 
\ion{He}{ii} $\lambda$4686 spectra in October 1989. This figure shows the stability of the underlying clock of $P_{\rm rot}$, and the similarity in behaviour between the lines in our ROI 
(dashed white lines) which probes the region just above the surface.
The spectra were averaged in bins of 0.025 and 0.05\,$P_{\rm rot}$ for the UV and optical spectra, respectively,  with the epochs indicated by arrows.  Two periods are displayed. Phase 0 corresponds to HJD $= 2451387.302 \pm 0.022$.
The vertical dashed black lines in the left two upper panels denote the rest wavelengths.  The lower left and middle panels are phased averages of the dynamic spectra presented in Fig.\,\ref{fig:xiSi4N4_87-94}. The fine structures 
in the \ion{Si}{iv} line 
are now smeared out by this averaging (in the left panel), highlighting its 
cyclic nature, whereas the pattern in the \ion{N}{IV} line shown in Fig.\,\ref{fig:xiSi4N4_87-94} appears quite similar supporting 
its strict periodicity 
(in the middle panel). 
{\textit{Left}:} 
ROI is indicated for both doublet lines, with the two according velocity scales at the bottom. The 
$\lambda$1402\,\AA\ component  was used in the analysis. The light grey maxima at phase 0.5 (and 1.5) clearly correspond to the dark maxima of the \ion{N}{IV} and \ion{He}{II} behaviour.   {\textit{Middle}:} Same but for \ion{N}{IV}. Note that the absorption appears first at intermediate velocity and decreases. {\textit{Right}:} Same but for \ion{He}{II}. 
The vertical dashed black lines are drawn at $\pm v_{\rm tot} = 230$\,km\,s$^{-1}$.
}
\label{fig:SiNHedynamic}
\end{center}
\end{figure*}
We should mention that the generalised Lomb-Scargle (GLS)  periodogram 
\citep{Ferraz-Mello:1981, Zechmeister:2009} 
did not show any pronounced peak at the 2.04061\,d period found with {\tt CLEAN}, unless the frequency grid included this period. 
This is illustrated in Fig.\,\ref{fig:LombScargleIUE}:  
the top panel shows the power based on a fine frequency grid, specified in period units by the range [0.54, 3.04]\,d  with step 0.001\,d, whereas the lower panel shows the same periodogram, but based on the range [0.54061, 3.04061]\,d, i.e.\ including the known period from {\tt CLEAN} even with a much coarser step of 0.01\,d. The first shows no indication of the known period, whereas the latter shows the peak within the error bar at the expected position, which exceeds the 99$\%$ confidence 
level by a factor $>5$. 
We note that the figure for the \ion{Si}{iv} data -- not shown -- is nearly identical. This critical dependency on the chosen frequency grid is obviously caused by the extremely irregular sampling of the 12\,yr dataset (Table\,\ref{tab:IUEtable}), and emphasises the need to use several period-search methods with care.

We used the values of the period, amplitude and phase provided by the {\tt CLEAN} algorithm as initial values for a weighted least-squares cosine fit  of the measured fluxes, $f$, in the ROI with four free parameters:

\begin{equation}
f(a,b,P,t_0,t) = b + a\, \cos\left(\frac{2\pi}{P}(t-t_0)\right),
\label{eq:Cosinefit}
\end{equation}
with $a$ the amplitude, $b$ the average flux, $P$ the period, and $t_0$ the epoch of maximum flux. The results are collected in Table\,\ref{tab:fitresults}.
The fit applied to the IUE \ion{N}{iv} fluxes gave a period of $2.040541 \pm  0.000034$\,d, with a reduced $\chi^2$ of 3.02.

To add the 11 STIS data points to the fit, several approaches are possible.  When adding the 11 individual STIS points with the $1/\sigma^2$ as weights to the IUE points, the formal solution of the best fit period becomes $P = 2.040473 \pm 0.000014$\,d (Table\,\ref{tab:fitresults}). 
Although the 1$\sigma$ error bars are much smaller than for the IUE data points, the very large range in S/N  (Table\,\ref{tab:STIStable}, depicted as dashed red circles in Fig.\,\ref{fig:STISNIVSIV}) is of obvious concern. 
With the S/N as weights, the fit would converge to a period close to its starting value, which indicates that no robust convergence can be reached. We therefore decided to group the 11 points into four groups. 
With these four fluxes and phases the resulting weighted fit gives $P = 2.040501 \pm 0.000023$\,d, i.e.\ within the error margins overlapping with the results of the IUE-only fit, which we therefore consider as a confirmation of the match between the absolute fluxes of the mean IUE and STIS spectra.

We now consider the \ion{Si}{IV} flux measurements. A least-squares weighted fit of the mean fluxes measured in the same  
ROI gave for the period 2.040546 $\pm$ 0.000037\,d, with a reduced  $\chi^2$ of 2.89, entirely consistent with the \ion{N}{iv} results. The difference in $t_0$  of 0.075\,d falls outside the 1$\sigma$ error bars of 0.023 (Table\,\ref{tab:fitresults}), indicating a likely phase shift.

We calculated the weighted mean of the IUE + 4w STIS \ion{N}{iv} and \ion{Si}{iv} periods, as listed in Table\,\ref{tab:fitresults}, which gives for the period 2.040514\,d with a formal 1-$\sigma$ uncertainty of 0.000020\,d. We adopt this value as the best period in Table\,\ref{tab:fitresults}. We calculated $t_0$ from a cosine fit of IUE + 4w STIS \ion{N}{iv} with the adopted period. This yields for the predicted epoch of maximum \ion{N}{iv} flux: 
\begin{equation}
{\rm \ion{N}{IV}\, \, HJD} = 2451387.302 \pm 0.022 + N \times (2.040514 \pm 0.000020),
\label{eq:maxphaseNIV}
\end{equation}
in which the first number was chosen near the middle of the first IUE and last STIS observation, and $N$ is an integer number.
(We note that in a preliminary report by \cite{sudnik:2018} the annotation `maximum' in their Fig.\,1 should have read `minimum'.)
The phase-folded \ion{N}{IV} and \ion{Si}{IV} fluxes are displayed in Figs.\,\ref{fig:NIVfit} and \ref{fig:SiIVfit}, respectively.
The best fits are overplotted. Note that the residuals, displayed in the lower panels, do not show evidence of an additional periodicity.

A visual confirmation is displayed in Fig.\,\ref{fig:SiNHedynamic}, which shows an overplot of all 307 phase-folded \ion{N}{IV} and \ion{Si}{IV} spectral lines, using quotient spectra from 
\cite{kaper:1996}. 
The emission part of the profiles did not vary significantly (see also Fig.\,\ref{fig:STISNIV}.), which is relevant as in strongly magnetic OB stars, variations in the shape of the P Cygni profile appears in both the absorption and emission components,
and are both modulated on the stellar rotation period (e.g. \citet{Henrichs:2011}, \citet{Erba:2021}).

We calculated the fractional phase of the individual velocity bins of the \ion{N}{iv}, \ion{Si}{iv} and \ion{He}{II} spectra for the adopted period, which can be compared  to Fig.\,15 in \cite{dejong:2001}. For a given line $i$ 
the fractional phase $\phi_i$  follows from 
\begin{equation}
\phi_i = \frac{t_{0_i}-t_0}{P},
\label{eq:fracphase}
\end{equation}
with $t_0$ and $P$ from Table\,\ref{tab:fitresults} (adopted values), and which was reduced to values between $-$1 and 1.  $t_{0_i}$ is the epoch of maximum flux in the individual velocity bin of the considered line. We calculated the error bars of the fractional phases as the propagated error according to Eq.~\ref{eq:fracphase}, using the formal errors listed in Table\,\ref{tab:fitresults}.

\begin{figure}[!t]
\begin{center}
\includegraphics[width=.75\linewidth]{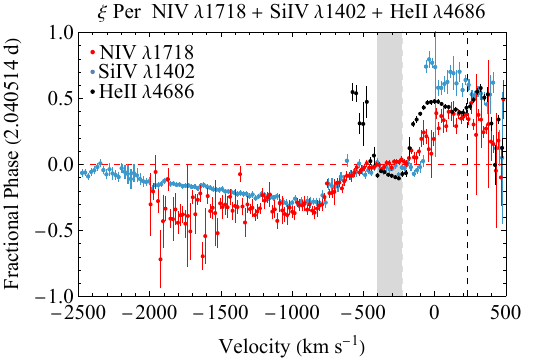}
\caption{Fractional phases from Eq.\,\ref{eq:fracphase} for the lines of \ion{N}{IV} (red), \ion{Si}{iv} (blue),  and \ion{He}{ii} (black) calculated in each velocity bin, while the period is kept constant. The relative flatness in the ROI (the shaded area) justifies these integration limits. The mean fractional phase of \ion{N}{IV} in this region is $0.00 \pm 0.01$. The vertical dashed lines are drawn at $\pm v_{\rm tot} = 230$\,km\,s$^{-1}$. Note the small phase delay for the \ion{Si}{iv} and \ion{He}{ii} mean values.}
\label{fig:fractionalphase}
\end{center}
\end{figure}

\begin{figure}[!ht]
\begin{center}
\includegraphics[width=.8\linewidth]{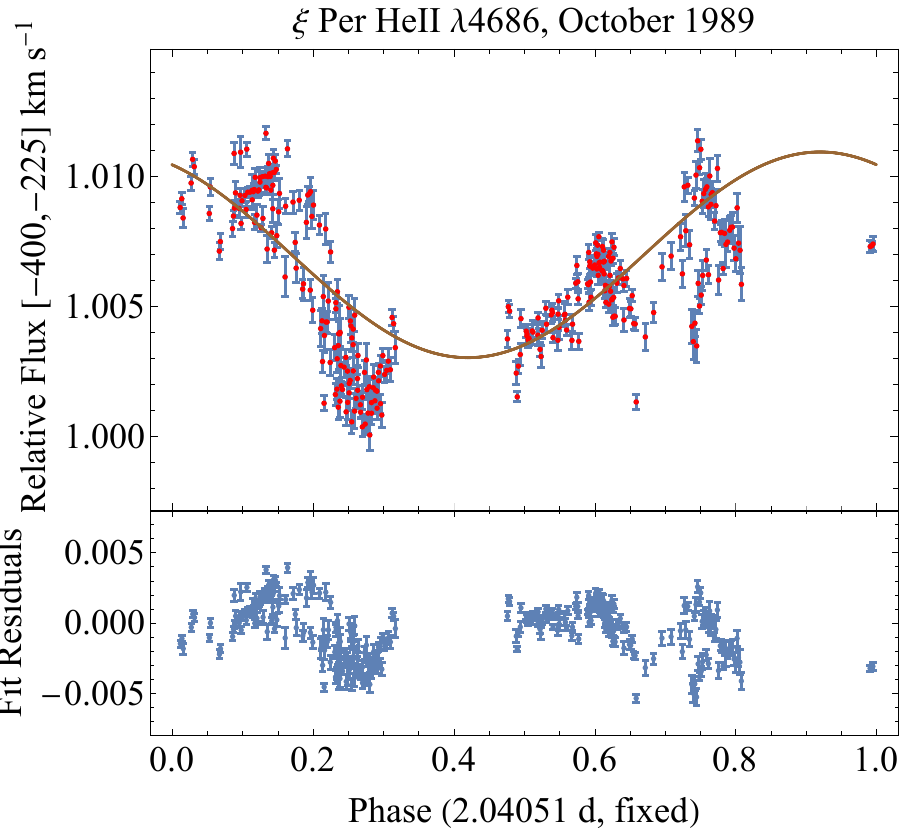}
\caption{{\textit{Top}:} Phase-folded relative flux measurements of the \ion{He}{II} $\lambda$4686\,\AA\ line in the interval [$-$400, $-$225]\,km\,s$^{-1}$ over 
5.5\,d. The best weighted cosine fit with a fixed period of 2.040514\,d is overplotted. Phase 0 corresponds to HJD\,$= 2451387.302 \pm 0.022$. {\textit{Bottom}:} Fit residuals.}
\label{fig:he2fit}
\end{center}
\end{figure}

\begin{figure}[ht!]
\begin{center}
\includegraphics[width=0.8\linewidth]
{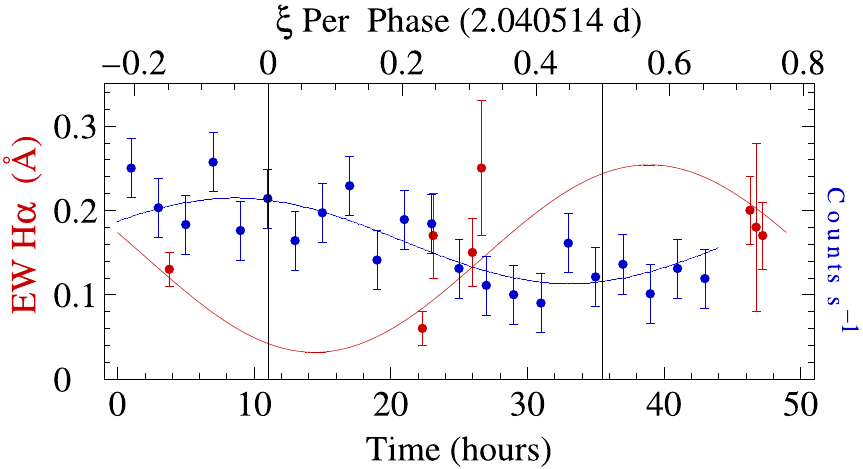}
\caption{
Revised version of Fig.\,8 by  \cite{massa:2014},  including \textit{Chandra} X-ray and H$\alpha$ data,
on the same horizontal scale 
converted to phase with our adopted period,  indicated on the top axis. The horizontal axis spans about one full rotation period and has its zero point at  HJD = 2453086.5904.
Red: Weighted cosine fit with a fixed 2.040514\,d period of H$\alpha$ EW data in Table 1 from \cite{massa:2014}.
Blue: \textit{Chandra} X-ray data given by \cite{massa:2014} with the cosine fit with our adapted period, overplotted on an arbitrarily adapted vertical scale. We see that maximum flux for both observables occurs 
near the same phase.
}
\label{fig:Chandra}
\end{center}
\end{figure}

The results are presented in Fig.\,\ref{fig:fractionalphase}.  The phase appears practically constant in the integration interval we used, which supports its validity. 
The small phase difference between \ion{N}{iv} and \ion{Si}{iv} in the ROI amounts to  $0.034 \pm 0.016$, as derived from the difference in $t_0$ of the fits with the fixed adopted period in Table \ref{tab:fitresults} divided by 
that period. This implies that the maximum flux in \ion{Si}{IV} occurs  earlier 
than in \ion{N}{iv}.

\subsection{\ion{He}{II} $\lambda$4686{\,\AA}}
\label{section:periodHe}

We folded the relative flux values of the \ion{He}{II} line with the adopted period, and applied a weighted fit of a cosine function with the remaining three free parameters. The results are presented in Table\,\ref{tab:fitresults} and Fig.\,\ref{fig:he2fit}, with a dynamic spectrum also included in Fig.\,\ref{fig:SiNHedynamic}. The maximum value occurs at 
\begin{equation}
{\rm \ion{He}{II}\, \, HJD} = 2447820.321 \pm 0.010 + N \times (2.040514 \pm 0.000020),
\label{eq:maxphaseHeII}
\end{equation}
in which we used the same format as in Eq.\,\ref{eq:maxphaseNIV}.

Although the scatter in the residuals is irregular and significant excursions do occur, the difference in maxima between the \ion{N}{IV} and \ion{He}{II} fits amounts to $ 1748.080 \pm 0.021$  cycles. i.e.\ very near an integer number, which is expected if the maxima are physically correlated.
The mean phase shift in the ROI is $-0.0834 \pm 0.0047$, noticeable in Fig.\,\ref{fig:fractionalphase}. See for a further discussion Sect.\,\ref{section:phases} and Fig.\,\ref{fig:phases}.

\subsection{Revisited H$\alpha$ - \textit{Chandra} co-variability}
\label{section:periodHa}

The co-variability of the H$\alpha$ line profiles with \textit{Chandra} X-ray fluxes  was   investigated by \cite{massa:2014}. However, there appears to be a mistake in their Fig.\,8, as their H$\alpha$ EW curve is incorrectly displayed with respect to the horizontal axis, which represents the \textit{Chandra} observations (A. Fullerton, priv.\ comm., {\citealt{Massa:2026}}),
although the epochs in their Table\,1 are correct.
A cosine fit to their H$\alpha$ EW data gives a maximum (or minimum flux) at HJD 2453088.212 $\pm$ 0.052, or 38.9 $\pm$ 1.3\,h after the beginning of the X-ray observations. A corrected version of their Fig.\,8 is given in Fig.\,\ref{fig:Chandra}. 
This maximum  perfectly  agrees with the maximum of the blue-wing H$\alpha$ EW presented by \cite{dejong:2001}, which had been obtained 1686 
cycles earlier, and is predicted to occur at $41.0 \pm 2.4$\,h.

\section{Interpretation of the period}
\label{section:rotation}

\begin{figure}[!t]
\begin{center}
\includegraphics[width=.7\linewidth]{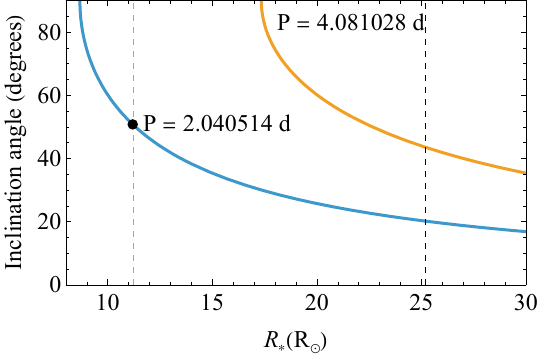}
\caption{Inclination angle as a function of stellar radius for the adopted rotation period (blue) and for twice this value (orange), based on $v \sin i$ $ = 215$\,km\,s$^{-1}$. The vertical dashed lines are drawn at the two possible solutions for the radius given by \cite{markova:2004}. A 4\,d period is excluded for a stellar radius smaller than 17.3\,$R_{\odot}$. The black dot is at the intersection point of the adopted stellar radius, implying an inclination angle of $51^{\circ}$.}
\label{fig:inclination}
\end{center}
\end{figure}

We identified the period of 2.040514\,d as the rotation period. An approximately two-times higher value had been previously argued by \cite{dejong:2001}, who used $R>17 R_{\odot}$, and by \cite{ramiaramanantsoa:2014}, who used $R=18.5 R_{\odot}$, the reason being that for such a large radius, the implied inclination angle for a 2-day period would be $i = 30^{\circ}$, which they found likely to be too small. They reasoned that at such a low inclination angle, the observed traveling bumps in atmospheric lines \citep{dejong:1999} are expected to go in both directions because a considerable part of the far surface beyond the pole would also be visible; whereas only one direction had primarily been observed. However, the much smaller radius of 11.2\,$R_{\odot}$ adopted in the current paper implies $i$\,=\,arcsin$ (P_{\rm rot}v \sin i/2\pi R) = 51 ^{\circ}$ (see Fig.\,\ref{fig:inclination}), which would accommodate much fewer red-to-blue traveling bumps. From this figure, we conclude that a 4~d rotation period is excluded for a stellar radius smaller than 17.3 $R_{\odot}$. We can exclude non-radial pulsations as being the underlying cause of these variations because the reported period of 3.5 h \citep{dejong:1999} in the \ion{He}{i} $\lambda$4713 line (derived from the same dataset used above) is far too short. If we assume that the strict 2 d periodic variation is caused by the presence of a magnetic dipole field, the fact that the fit of the 
flux variations
show only one single sine wave, with no 
evidence of a double sine wave (Fig.\,\ref{fig:NIVSiIVfit}), implies that only one pole is visible during a full rotation. The magnetic axis should therefore make an angle with the rotation axis of $\beta \leq 90^{\circ} - i = 39^{\circ}$ for a centred dipole.
The actual values of $i$ and $\beta$ depend of course on the adopted value for the stellar radius, the uncertainty of which is determined by the uncertainties of the distance and the angular diameter. 
We also attempted to fit a double sine wave with periods $P_{\rm rot}$ and $P_{\rm rot}/2$ to the \ion{N}{IV} flux data, similar to the approach by \cite{neiner:2003b}, but this did not lead to a better result.

One crucial thing to note is that the rotation period that we detect appears to fulfill both roles mentioned in Sect.\,\ref{sect:intro}. Namely, for the first time in an OB star, we find that it provides a stable clock that subtends strictly periodic variations in the wind near the stellar surface, as illustrated in Fig.\,\ref{fig:SiNHedynamic}. However, as clearly shown in Fig.\,\ref{fig:xiSi4N4_87-94}, it is also associated with more cyclic variable behaviour, consistent with the well-known DAC phenomenology. We suggest that this points to the existence of {two distinct drivers:} 
one that must remain stable over timescales of years to decades (i.e. for the entire temporal span of $\xi$\,Per's UV dataset), and one that varies in a more stochastic fashion, in a manner consistent with the driving of DACs in other OB stars (presumably spots or surface inhomogeneities whose physical origin remains to be conclusively determined, as previously mentioned). For the former, the only mechanism known to trace rotation over long periods of time is a fossil, large-scale magnetic field.

\section{Space photometry}
\label{Section:photometry}

In addition to the spectroscopic datasets presented above, $\xi$\,Per  was  observed by three high-precision space-based broadband optical photometers: the Microvariability and Oscillations of STars (MOST) mission \citep{Walker:2003}, the BRIght Target Explorer (BRITE; \citealt{Weiss:2014}), and the Transiting Exoplanet Survey Satellite (TESS; \citealt{Ricker:2015}). Together, these observations  span about 
13 years between 2011 and 2024.

\subsection{{\textit{MOST}}}

\cite{ramiaramanantsoa:2014} presented  four weeks of continuous 
photometry with a cadence of 20\,min obtained with the MOST satellite in November 2011. 
The dynamic plot of the periodogram in their Fig.\,3 shows a pronounced peak starting at about 2.15\,d, which wiggles irregularly to about 1.8\,d at the end $-$ never remaining at our adopted rotation period. This period was also not found between the 12 most significant periods from a global period analysis. The light curve was modelled with several corotating bright-spot configurations on the star with a 4.18\,d rotation period. Random starting times and lifetimes vary from one spot to another. They concluded that this was the first convincing reported case of corotating bright spots on a O star.\\

\subsection{{BRITE}}

\begin{figure}[!ht]
\begin{center}
\includegraphics[width=.8\linewidth]{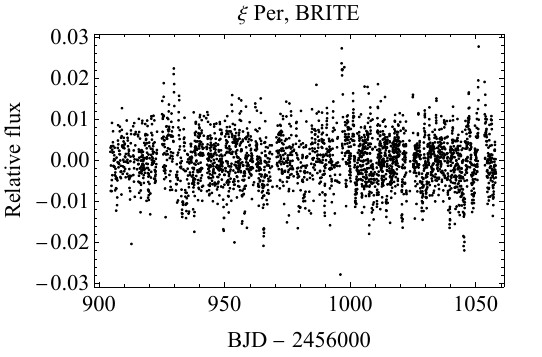}
\caption{BRITE light curve 2014 November till February 2015.}
\label{fig:BRITElightcurve}
\end{center}
\end{figure}

\begin{figure}[!ht]
\begin{center}
\includegraphics[width=.8\linewidth]{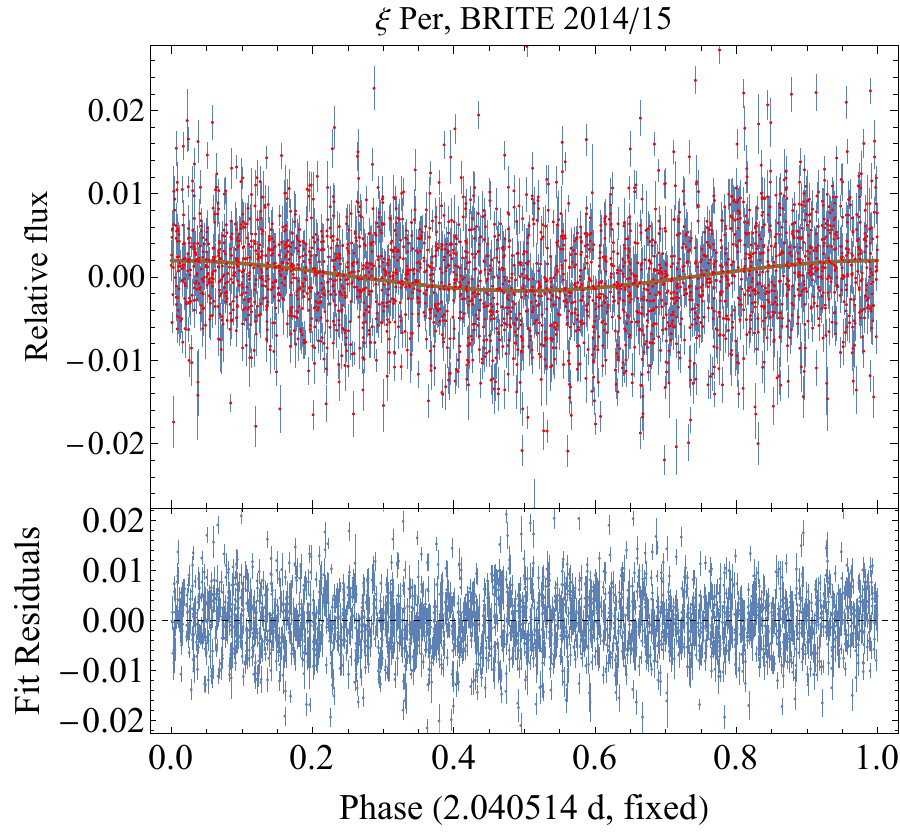}
\caption{\textit{Top}:  Folded light curve of the BRITE 2014/15 data with the rotation period.
Phase 0 corresponds to BJD = 2456982.842 $\pm$ 0.027 (
Eq.\,\ref{eq:maxphaseBRITE}). 
The best cosine fit is overplotted in brown. {\textit{Bottom}:} Fit residuals in the same units as the top panel. The large scatter is considered to be intrinsic. 
}
\label{fig:BRIGHTfit}
\end{center}
\end{figure}

\begin{figure}[!ht]
\begin{center}
\includegraphics[width=0.7\linewidth]
{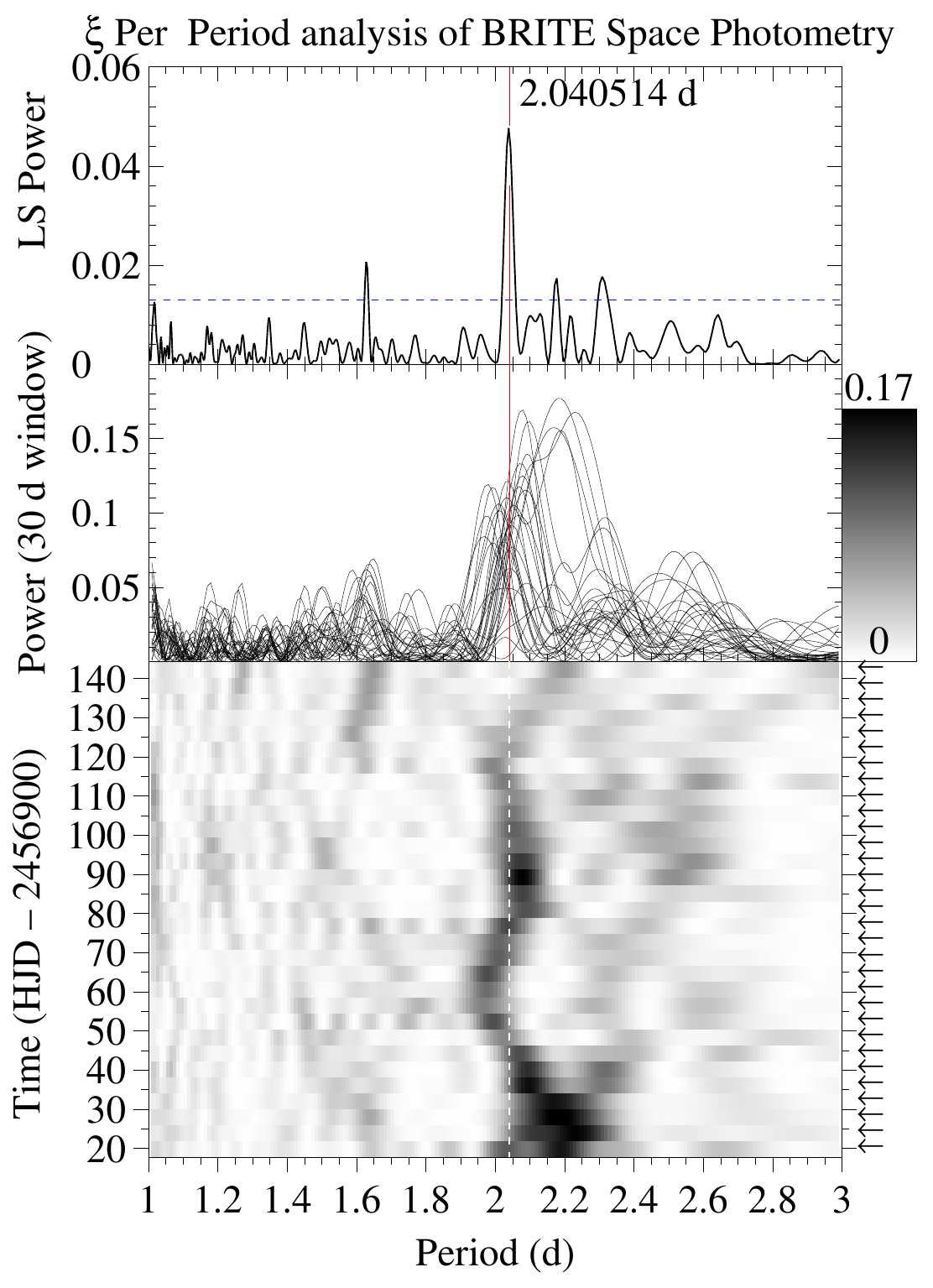}
\caption{\textit{Top}:  
Lomb-Scargle periodograms derived from the BRITE light curve. 
The peak at 2.039 $\pm$ 0.015\,d is consistent with $P_{\rm rot}$ 
(vertical red line), exceeding 3.7 times the 99$\%$ confidence level (horizontal dashed blue line). 
\textit{Middle}: Superposed periodograms of 30\,d wide sliding windows, with a step equal to 2$P_{\rm rot}$.
\textit{Bottom}: Grey-scale representation of the middle panel, with the central epochs of the windows indicated by arrows. 
The wiggles around the adopted period clearly show that a subset of 30\,d will unlikely reveal the global period. 
}
\label{fig:periodogram}
\end{center}
\end{figure}

The BRITE satellite consortium \citep{Weiss:2021} observed $\xi$\,Per at two extended periods: during 153 days starting at 4 September 2014 (referred to as BRITE14/15) and  during 176 days starting at 14 September 2016. The de-correlated light curve of the first dataset is available for download from the BRITE public archive\footnote{\url{https://brite.camk.edu.pl/pub/index.html}}, and is, after conversion to relative flux, shown in Fig.\,\ref{fig:BRITElightcurve}. The second dataset will be analysed 
in a separate paper (Ramiaramanantsoa, priv.\ comm.). 

\begin{figure*}[!t]
  \centering
    \subfigure[]{\label{fig:TESSlc59}
    \includegraphics[width=0.32\textwidth]{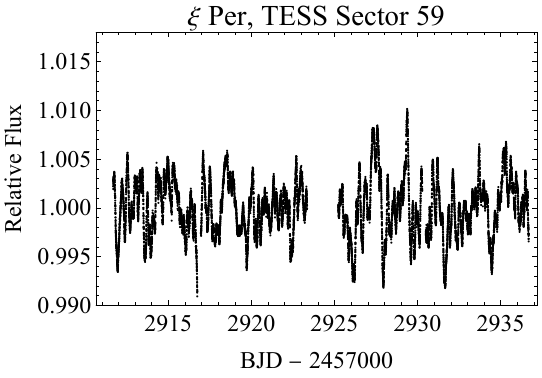}}
    \subfigure[]{\label{fig:LSxiPerTESSlc86}
    \includegraphics[width=0.32\textwidth]{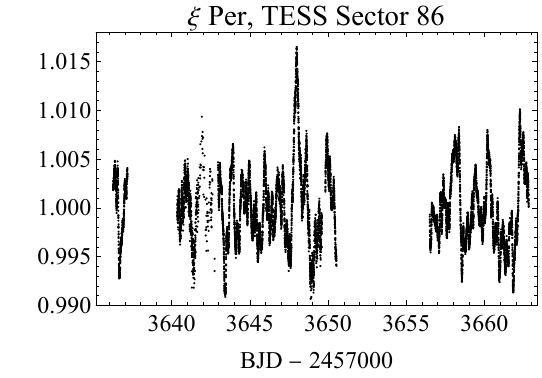}}    
    \subfigure[]{\label{fig:TESSLS}
    \includegraphics[width=0.32\textwidth]{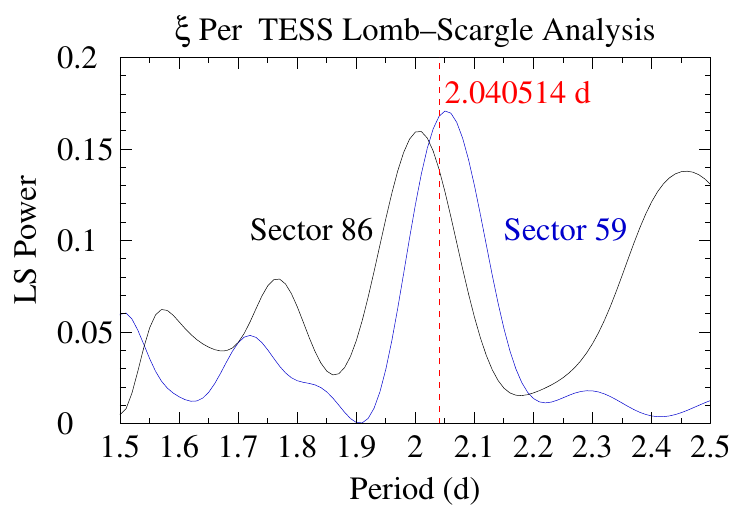}}
\caption{
\textit{Panel (a)}: TESS 
light curve from 2022 November 26 until
December 23. 
\textit{Panel (b)}: Light curve from 2024 July 27 until August 23.
\textit{Panel (c)}: Lomb-Scargle periodograms for Sectors 59 and 86 of TESS data. The vertical dashed red line indicates our adopted rotation period. The similarity with the middle panel of Fig.\,\ref{fig:periodogram} is clear: the peak wanders around the adopted period.
}
\label{fig:TESS5686LS}
\end{figure*}

\cite{Zwintz:2024} noted `clear variability with a frequency of 0.49\,d$^{-1}$' in the first dataset, without details. 
A Lomb-Scargle period analysis 
yielded a very pronounced peak at 2.039  $\pm$ 0.015\,d (Fig.\,\ref{fig:periodogram}, top), 
confirming the note by \cite{Zwintz:2024}, which
exceeds  the 99$\%$ confidence 
level by a factor 3.7, as determined by the bootstrap method 
calculated by comparison with white noise, although this may not applicable for this dataset, as the light curve shows 
large intrinsic non-periodic excursions. 
A weighted cosine fit (Eq.\,\ref{eq:Cosinefit}) with all four parameters free yields $P = 2.0389 \pm 0.0013$\,d, which differs only by 0.0016\,d from our adopted period. 
A similar fit but with the period parameter fixed at $P_{\rm rot}$ gave essentially the same result, except that 
$t_0$, the epoch of maximum light shifted 0.003\,d, well within the 
$\sigma_{t_0}\,=\,0.027$\,d error bar. We therefore adopted this
last fit (Table\,\ref{tab:parameters})
yielding for the maximum flux to occur at 
\begin{equation}
{\rm BRITE\, BJD} = 2456982.842 \pm 0.027 + N \times (2.040514 \pm 0.000020).
\label{eq:maxphaseBRITE}
\end{equation}

\noindent The maximum flux with respect to the adopted phase of maximum flux derived from the \ion{N}{IV} spectra (Table \ref{tab:fitresults}) occurs 
at  $0.45 \pm 0.05$\,d, or in phase: $\phi_{\rm BRITE max} = 0.22 \pm 0.03 $ (Fig.\,\ref{fig:phases}).

The phase-folded light curve, with the overplotted best fit with the period fixed at $P_{\rm rot}$, together with the fit residuals is presented in Fig.\,\ref{fig:BRIGHTfit}. It is clear that the scatter around the fitted curve is very large, roughly 3 times the amplitude, which is also apparent from the reduced $\chi^2$ = 11. 
We accept the periodicity because 
(1) there is only one dominant peak in the one magnetic pole that we do see is
at its closest alignment to the line of sight the periodogram, (2) this period coincides within its error bar with $P_{\rm rot}$, 
and (3) this peak exceeds the estimated 99$\%$ confidence level by a factor  $> 3$, although see the caveat described above. 
To further investigate the contrast with the earlier MOST results \citep{ramiaramanantsoa:2014}, which covered only 29 days, we calculated the periodograms in windows of 
31\,d, each containing about 100 data points, covering 15 rotations. We slided the window throughout the full 153\,d observing time, in steps equal to 2\,$P_{\rm rot}$.
The middle panel of Fig.\,\ref{fig:periodogram} shows the overplot of the resulting periodograms, 
with the dynamic evolution in the lower panel. 
We see that the maximum peak wiggles irregularly about 10$\%$ on typical timescales varying from 10 to about 40\,d, and coincides only in a few windows with $P_{\rm rot}$. 
There are also windows with a less 
significant periodicity, for example near the end of the 153\,d global observing window. This behaviour is 
similar to 
the absence of a clear periodicity in the MOST light curve (Fig.\,3 in \citealt{ramiaramanantsoa:2014}). 
We conclude that the MOST and BRITE light curves show similar characteristics, with a global small-amplitude (2\,mmag) component which is only apparent if a sufficiently long ($> 20 P_{\rm rot}$) stretch of data is considered. The variations are dominated by a superposed 
randomly distributed large scatter with 10\,mmag amplitude. 
This is consistent with a more stochastic component such as that which might be expected from the drivers of DACs, causing large cyclic variations superposed on top of a low-level periodic signal due to a weak, large-scale magnetic field. Considering the large hourly 
fluctuations we also searched for significant power around the 4.45\,h periodicity, attributed by \cite{dejong:1999} as due to non-radial pulsations, but did not find any.

Ramiaramanantsoa (priv.\ comm.) found a peak around  2.0467\,d in the {\tt CLEAN}ed power spectrum, close to the value derived above, with an approximate sine-wave shaped light curve with a peak-to-peak amplitude of $\sim$\,5\,mmag. A combined period analysis of both BRITE data sets would be very useful to further characterise the photometric behaviour, which will be the subject of a separate study. 

\subsection{{TESS}}

{TESS} observed $\xi$\,Per = TIC 94614234 with a 2 minute cadence during sector 59 from 2022 November 26 until
December 23, and during sector 86 from 2024 July 27 until August 23. The detrended and 
normalised light curves obtained using  {\tt Lightkurve}  \citep{LightKurve:2018L}  are presented in Fig.\,\ref{fig:TESS5686LS}.
Typical uncertainties are 0.05$\%$ of the flux. We note that the TESS and  BRITE photometric variations are comparable (Fig.\,\ref{fig:BRITElightcurve}), both showing occasional spikes. The corresponding standard Lomb-Scargle periodograms are displayed in Fig.\,\ref{fig:TESS5686LS} (right), 
showing a very pronounced maximum at $2.050 \pm 0.070$\,d for Sector 56, and $2.011 \pm 0.078$\,d for Sector 86,  with the uncertainties being the half width at half maximum. These values overlap our adopted 
period, as indicated by the vertical dashed red line. 
It is clear that the relatively short observing window of 27.4 days does not allow for a very accurate determination of the period, as compared to the five times longer BRITE window. The deviations from the adopted period is strikingly 
similar to the BRITE results, as shown in the middle panel of Fig.\,\ref{fig:periodogram}. 
The similarity between the TESS and BRITE light curves regarding the very significant hourly fluctuations poses the same problem to quantify a confidence level as described above, which we therefore did not attempt.
As in the BRITE periodogram, we did not find any significant power around the 4.45\,h periodicity, attributed by \cite{dejong:1999} as due to non-radial pulsations.

\subsection{Summary of phases}
\label{section:phases}

\begin{figure}[!ht]
\begin{center}
\includegraphics[width=0.75\linewidth]{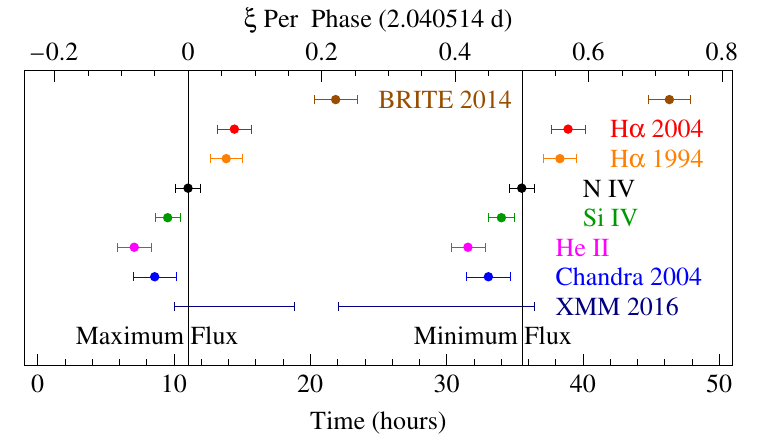}
\caption{
Phases of maximum and minimum flux, plotted on the same horizontal scale of the 2004 campaign including 
the \textit{Chandra} X-ray data, spanning
about one full rotation period, with
its zero point at HJD = 2453086.5904. 
The flux extrema of the fit to the H$\alpha$ data from this campaign 
(red) overlap the extrapolated epochs from \citet[orange]{dejong:1999}. 
The error bars are calculated from the width of the extrema as marked in their Figs.\,4 and 5, and using the propagated error. 
The reported ranges of maximum and minimum  XMM X-ray flux \citep[dark blue;][]{massa:2019} are consistent with the epochs of the maximum X-ray flux of \textit{Chandra} 
and with the predicted epochs of the \ion{N}{IV} (black), \ion{Si}{IV} (green), and \ion{He}{ii} (magenta) flux maxima according to Eq.\,\ref{eq:maxphaseNIV}. The epochs of the minimum X-ray flux of \textit{Chandra} and XMM are also formally consistent, and marginally consistent with the epoch of predicted \ion{N}{IV} flux minimum.
Note: the significant trend in phase, showing the progress in time of \ion{He}{ii}, \ion{Si}{iv}, \ion{N}{iv} (as  expanded in Fig.\,\ref{fig:fractionalphase}) and H$\alpha$.
The epochs of the maximum and minimum photometric flux of the BRITE14/15 campaign are indicated in brown, and show 
a phase delay of $\approx 0.2$ with respect to the spectral line fluxes. 
}
\label{fig:phases}
\end{center}
\end{figure}

A summary of the relative phases of maximum and minimum flux of H$\alpha$, \ion{N}{iv}, \ion{Si}{iv}, \ion{He}{ii} in the
ROI is presented in Fig.\,\ref{fig:phases}, with their respective error bars propagated to the epoch of the broadband X-ray observations by \textit{Chandra}. The upper horizontal axis shows the \ion{N}{iv} phase (Eq.\,\ref{eq:maxphaseNIV}). The epochs of maximum and minimum \textit{Chandra} X-ray fluxes of the 
cosine fit are  indicated. The (broad) intervals of maximum and minimum flux of the XMM observations in 2016 as determined by \citet{massa:2019} are also marked. We note that these intervals are wavelength dependent. The \textit{Chandra} and XMM extrema are more or less overlapping. 
It should be noted that Fig.\,\ref{fig:phases} shows a significant order in phase of maximum flux:  \ion{He}{ii}  - \ion{Si}{iv} - \ion{N}{IV}, followed by H$\alpha$. The 
formation regions of these spectral lines are clearly related. The relative phases of the X-ray flux maxima are less clear.  The 
deviating phase of BRITE maximum clearly points to a different origin. This is discussed in Sect.\,\ref{section:DACs}.

\section{Magnetic measurements}
\label{section:magnetic}

\begin{figure}[!t]
\begin{center}
\includegraphics[width=.7\linewidth]{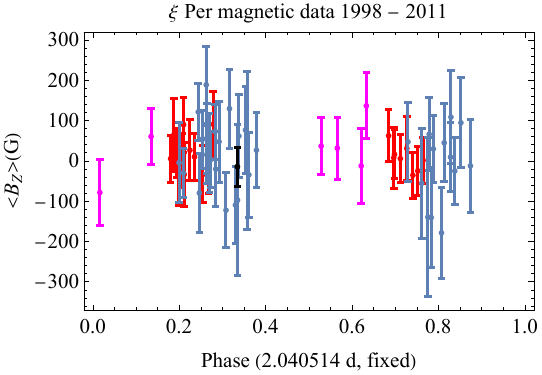}
\caption{Magnetic measurements $\langle B_{Z}\rangle$ of $\xi\,$Per phase-folded with our adopted period based on Eq.\,\ref{eq:maxphaseNIV}, i.e.\ maximum flux at phase 0.  Magenta: H$\beta$ UWO at CFHT 1994.  Blue: MUSICOS at TBL 1998 (points with error bars $> 200$\,G omitted).  Red:  NARVAL at TBL 2006 and 2007.  Black: ESPaDOns at CFHT 2011.}
\label{fig:B}
\end{center}
\end{figure}

Magnetic measurements $\langle B_{Z}\rangle$ of $\xi\,$Per  were obtained at the 2m T\'{e}lescope {\it Bernard Lyot} (TBL) at the Pic du Midi Observatory, France) and at the 3.6m Canada-France-Hawaii Telescope (CFHT) on Hawaii. At TBL the spectropolarimeters MUSICOS (in 1998) and NARVAL (in 2006 and 2007)  were  used \citep{Donati:1999}, whereas the acquisition at CFHT was done with the UWO H$\beta$ (in Oct 1994) and ESPaDOnS (in 2011) spectropolarimeters \citep{Donati:2006c}. We collected the data from \cite{dejong:2001} and \cite{david-uraz:2014}, where $\langle B_{Z}\rangle$ denotes the longitudinal component of the magnetic field in the line of sight integrated over the stellar surface. Although the technique 
to measure $\langle B_{Z}\rangle$ in these two papers differ, 
we still plotted the  phase-folded magnetic data together in Fig.\,\ref{fig:B}.  
We omitted 16 MUSICOS data points that had error bars exceeding $200$\,G. 

Most points appear to 
cluster together to the extent that the data become sparse around phases 0 and 0.5, 
where the extrema of the longitudinal magnetic field curve are expected 
in the case of a 
large-scale dipolar field. 
In particular, phase 0 is where we would expect a magnetic pole to point closest to the line-of-sight, leading to a maximum 
in |$\langle B_{Z}\rangle$| 
and, hence, the best chance 
of a magnetic detection.
We note that around these important phases only a few UWO data points from 1994 based on only the H$\beta$ line are available, which are supposedly characterised (in spite of their similar formal error bars) by a lower degree of accuracy than the measurements based on the least-square deconvolution technique
applied to ESPaDOns spectropolarimetry in 2011 at the same telescope. 
New magnetic measurements, taken at the appropriate phases, might constrain the field strength better than the present upper limit of 59\,G on the polar strength as derived by \cite{david-uraz:2014} based on the NARVAL and ESPaDOns data. 

Additionally, the Bayesian analysis (based on the method of \citealt{petit:2012}) that was performed by \citet{david-uraz:2014} assigned arbitrary phases to each observation, which means that using a fixed period would also yield tighter constraints on the dipolar field strength. Finally, it should be noted that a reanalysis of the same data by \cite{grunhut:2017}, but using regularisation when computing the least-square deconvolution profiles, yielded lower uncertainties on the longitudinal field measurements (by about one half), also pointing to a possible lower upper limit.

\section{Discussion}
\label{section:discussion}

\subsection{Simultaneous UV and H$\alpha$ spectroscopy}
\label{subsection:UVHa}
The most comprehensive multiwavelength study of $\xi\,$Per  was presented by \cite{dejong:2001}, which included UV, H$\alpha$, \ion{He}{I} $\lambda$6678 \,\AA, \ion{He}{I} $\lambda$6683 \,\AA, and spectropolarimetric measurements during 10 consecutive days in October 1994. One 
main conclusion was that the blue wing of the H$\alpha$ line varied in concert with the \ion{N}{IV} low-velocity part, more or less similar to the correlation we found above for the \ion{He}{II} line. They used a rotation period of 4.18\,d based 
on a stellar radius of at least 17\,$R_{\odot}$, 
from \cite{puls:1996}, although they mention that 12\,$R_{\odot}$ derived by \cite{leitherer:1988} could not be excluded,
in agreement with  
\cite{dejong:1999}. The very dense simultaneous coverage of H$\alpha$ and IUE spectroscopy in \cite{dejong:2001} 
yielded  
the epoch HJD $2449645.84 \pm 0.05$ for 
the maximum absorption in the blue wing of H$\alpha$ 
(Table\,\ref{tab:fitresults}), more than 1000 rotation cycles later than the centre of the IUE \ion{N}{IV} observations. 
A comparison with the predicted maximum absorption in \ion{N}{IV} (Eq.\,\ref{eq:maxphaseNIV}) gives a phase difference of $0.04 \pm 0.05$, i.e.\ consistent with zero, 
confirming that these spectral lines are formed in the same region. The synchronous behaviour of the DACs in the UV resonance lines at higher velocity and the low-velocity regions  
is well documented in \cite{dejong:2001}.  The phase differences are $0.92 \pm 0.03$, $0.06 \pm 0.04$, and $0.05 \pm 0.04$ for the \ion{Si}{IV} $[-2500, -200]$\,km\,s$^{-1}$, \ion{C}{IV} $[-1100, -200]$\,km\,s$^{-1}$, and \ion{N}{V} $[-1100, -200]$\,km\,s$^{-1}$ lines, respectively.
They noticed 
that over the course of 10 days a strong DAC appeared every 2 days, 
always preceded by a weaker DAC, which peaked 
about one day earlier, as 
also observed by 
\cite{kaper:1999}. In this work, we interpret 
these strong cyclic DACs being 
caused by CIRs driven by stochastic surface spots, while we understand the weaker DACs 
to in fact not be DACs at all but instead periodic in nature and 
putatively 
caused by a large-scale magnetic field. 

It is also quite informative to consider phasing comparisons between UV line profile variations and H$\alpha$ emission in known magnetic O-type stars. In that context, the phase corresponding to maximum H$\alpha$ emission is typically referred to as the high state, with the phase of lowest emission being labeled as the low state. As shown in several stars (e.g. \citealt{marcolino:2012,marcolino:2013,naze:2015}) and summarised by \citet{David-Uraz:2019a} -- in particular as shown in their Fig.\,3 -- while the overall line profile variations are quite complex and not obviously monotonic across the whole velocity range of the UV resonance lines, the blueward fluxes at low velocity in both \ion{Si}{IV} and \ion{C}{IV} lines appear to generally be at their highest at H$\alpha$ high state, a behaviour that is consistent with what we see in the \ion{N}{IV} line of $\xi\,$Per.

\subsection{Simultaneous X-ray observations and H$\alpha$ and UV
spectroscopy}

\cite{massa:2014} concluded from a cosine fit that the X-ray minimum occurred at $32.9 \pm 1.7$\,h since the start of the observations.
They argued that the rather close coincidence of this minimum and H$\alpha$ absorption maximum suggests that the X-ray and H$\alpha$ emission is obscured by the same material. 
In Sect.\,\ref{section:periodHa} we 
concluded instead that the X-ray minimum occurred $5.8 \pm 2.9$\,h prior to the H$\alpha$ absorption maximum, 
or $0.12 \pm 0.06$ prior in phase. 
We reproduced their Fig.\,7 after correction  
in our Fig.\,\ref{fig:Chandra}. 
The expected maximum \ion{N}{IV} flux on this scale is at 11.02 $\pm$ 0.40\,h
(Eq.\,\ref{eq:maxphaseNIV} with $N$ = 833). 
These conclusions, however,  
depend on the acceptance of the cosine fit. The hardly less significant
linear fit would not allow for such a conclusion.

We note that the X-ray spectral line analysis and subsequent H$\alpha$ modelling of \textit{Chandra} spectra of $\xi$\,Per by Cohen et al. (2014) showed that the X-ray
emission region starts around 1.5\,$R_{\ast}$.
This modelling 
was predicated upon the method of \citet{owocki:2001}, 
in terms 
of embedded wind shocks (e.g. \citealt{feldmeier:1997}) in a wind treated as essentially spherically symmetric (with the exception of small-scale structure). However, the geometry of a magnetised stellar wind is not spherically symmetric, and strong, X-ray-producing shocks can occur closer to the stellar surface due to wind material launched from both magnetic hemispheres colliding almost head-on at the magnetic equator (the so-called magnetically confined wind shock scenario; \citealt{babel:1997}). The applicability of either of these models to the X-ray analysis of $\xi\,$Per could be further constrained using other diagnostics, such as the $f/i$ ratio of He-like triplets (e.g. \citealt{leutenegger:2006}). Such an 
analysis is, however, beyond the scope of this paper.

The parallel behaviours of H$\alpha$ and X-ray fluxes is also seen in other magnetic O-type stars, e.g.\
the strongly magnetic O7\,Vp star $\theta^1$\,Ori\,C, 
with a magnetically confined wind 
with a disc around its magnetic equator \citep{donati:2002}, causing its maximum X-ray emission to coincide more or less in phase with maximum H$\alpha$ emission, 
when the northern magnetic pole is closest to the observer's line of sight 
\citep[][their Fig.\,4]{gagne:2005}. 
Similarly, \citet{petit:2015} describe a very analogous parallel behaviour of both X-ray and H$\alpha$ fluxes for the most strongly magnetised known O star, \object{NGC 1624-2}, and offer a schematic view.  
The flux variations can be understood in terms of 
rotational modulation of 
the projected area of this 
magnetospheric emitting region.  
A similar behaviour is seen for HD\,191612, with magnetohydrodynamic (MHD) simulations seemingly validating this overall phenomenological picture \citep{naze:2016}. 

\cite{massa:2019} presented an analysis of 11 HST STIS spectra. They concluded that the 
\ion{N}{iv} and \ion{Si}{iv} variations during the four days in 2016 are shifted in phase 
while our analysis shows that these are strongly
correlated. 
In Sect.\,\ref{section:STIS} we argue that the absolute flux values in the \ion{Si}{iv} region are unreliable, and hence we dispute their conclusion of a phase shift.

\subsection{Wind magnetic confinement}

\begin{figure*}[ht!]
\centering
\includegraphics[width=0.9\linewidth]{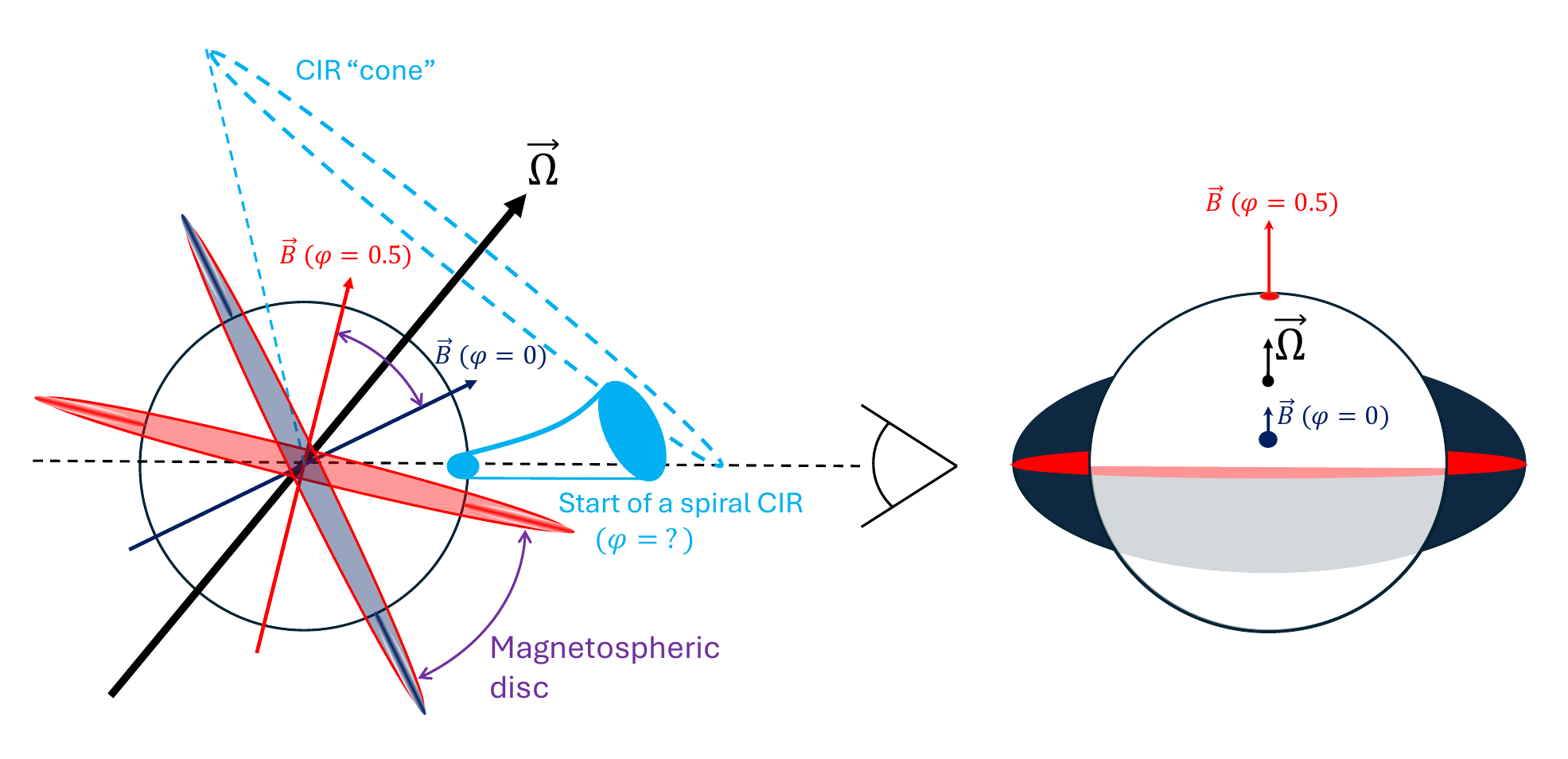}
\caption{
Schematics summarising our proposed theoretical framework, allowing for both periodic and cyclical variability, with the former 
being caused by the varying projected area of the magnetospheric disc due to rotation. \textit{Left:} Schematic showing the suggested physical drivers of both the periodic variability seen at low velocity in the \ion{N}{IV} 
line (as in Fig.\,\ref{fig:xiSi4N4_87-94} (bottom), 
believed to be caused by magnetically confined wind material close to the stellar surface, shown in red at the so-called low state of emission, and in dark blue at the high state) and more traditional, cyclical DACs in \ion{Si}{iv} (as in Fig.\,\ref{fig:xiSi4N4_87-94} (top), believed to be caused by coexisting
bright surface 
spots, driving CIRs, an example of which is shown in light blue). 
The observer is
placed at infinity 
on the right. 
For viewing purposes, we use 
$i$ = $51^\circ$ 
and $\beta = 25^\circ$. 
The magnetospheric region 
believed to be responsible for extra emission in H$\alpha$, X-rays, as well as in low-velocity regions of UV lines, is shown in dark red or dark blue, depending on the rotational phase. The `CIR cone', i.e.\ 
the cone on which the CIR extends from the surface, is shown in dashed light blue  
curves 
-- for a 3D example 
see Fig.\,7 by \citet{dessart:2004}.
\textit{Right}: Similar schematic, but seen from the point of view of the observer at phases of maximum and minimum flux (the CIR is removed from this view for clarity). 
The phase of the photometric flux minimum, which trails the minimum flux in the lines by $\approx 0.2$ in phase (Fig.\,\ref{fig:phases}) might be caused by obscuration effects by the magnetospheric disc and the CIRs.}
\label{fig:schematic}
\end{figure*}

If, for the sake of argument, we assume that $\xi$ Per hosts a magnetic dipole field, its effect on the wind flow can be characterised by the wind magnetic confinement parameter, introduced by \cite{uddoula:2002}, defined as $\eta_{*} = B^2_{\rm eq}R^2/\dot{M}v_{\infty}$, where $B_\textrm{eq}$ is the surface equatorial magnetic field strength, equal to half of the surface polar strength, $B_\textrm{p}$, for a centred dipole, $R$ is the stellar radius, $\dot{M}$ is the wind mass-loss rate, and $v_\infty$ is the wind terminal velocity. This parameter measures the ratio of the magnetic to wind kinetic energy density. For parameters from Table\,\ref{tab:parameters} and an upper limit on the polar field strength of 59\,G \citep{david-uraz:2014}, which corresponds to $B_{\rm eq}=B_{\rm p}/2=29.5\,$G, we obtain $\eta_{*} = 0.16$. 
\cite{uddoula:2002} showed that even for $\eta_{*} = 0.1$ significant deviations of spherical outflow will occur which can divert wind material from higher magnetic latitudes towards the magnetic equator, although the deviations from spherical symmetry are largely confined to the region near the magnetic equator, with the overall wind structure at higher magnetic latitudes being largely unaffected by the field. For $\xi$\,Per, $\eta_* = 0.1$ corresponds to a magnetic field strength of $B_{\rm eq}\approx$ 23\,G, below the upper limit reported by \citet{david-uraz:2014}. Further, it should also be noted that the weak confinement limit has been scarcely explored with MHD simulations, such that there is no consensus threshold value of $\eta_*$ below which a magnetic field would be considered as dynamically insignificant for a massive star's wind. It is clear that the current magnetic measurements depicted in Fig.\,\ref{fig:B} could well accommodate a dipolar field corresponding to values of $\eta_{*}$ in a range around 0.1.

With our adopted stellar parameters and an assumed value of $B_{\rm eq} = 29.5$\,G
the Alfv\'{e}n radius, where the flow velocity equals the Alfv\'{e}n speed, is located at  \citep{petit:2013} $R_{\rm A}/R = 0.3 + (\eta_{*}+0.25)^{1/4}=1.1$ for $\eta_{*} = 0.16$
and smaller for smaller values of $\eta_{*}$, which is well inside the Keplerian corotation radius of $R_{\rm K}/R \equiv (GM/(2\pi/P)^2)^{1/3} = 1.8$, at which the centrifugal
force balances the gravity. This implies that $\xi$ Per is in this sense a slow rotator like most O-stars, and would host a mini dynamical magnetosphere if such a field is present, as opposed to
rapidly rotating stars with strong magnetic fields, which carry a centrifugal magnetosphere \citep{petit:2013}.

If we assume $i = 51^{\circ}$ and $\beta < 39^{\circ}$ (the latter value as a consequence of hiding one of the two poles, cf.\ Sect.\,\ref{section:rotation}), we would expect for the ratio of the longitudinal
field minimum to the longitudinal field maximum \citep{preston:1967} equal to \mbox{cos($\beta + i)$/cos($\beta - i) \approx 0$}, or very small for $\beta = 39^{\circ}$, but about 0.17 for $\beta \approx 30^{\circ}$ (i.e. rather sensitive to the obliquity angle).

\subsection{On the origin of 
ultraviolet line profile variability in $\xi$ Per}
\label{section:DACs}

In this subsection we attempt to sketch a consistent picture to account for both periodic and cyclic UV line profile variations in $\xi$\,Per assuming   
a centred 
oblique magnetic dipole. 
Based on the similar phasing between fluxes in H$\alpha$, low-velocities in \ion{N}{iv}, 
\ion{Si}{IV}, and X-rays, as well as similar behaviour observed in several magnetic early-type stars, we ascribe the periodically varying fluxes in all three of these diagnostics to magnetically-confined wind material found close to the stellar surface, mostly along a disc of magnetospheric material concentrated near the magnetic equator. As the star rotates, the misalignment between the rotation and magnetic axes (which, as argued above, is expected to be $\beta < 39^\circ$) gives rise to a changing projected area of this disc for the observer.
For smaller angles between the magnetic axis and the line of sight, this disc is off the limb of the star and contributes added emission, whereas for large angles, the disc is seen nearly edge-on and contributes added absorption by obscuring the stellar surface, as well as reduced emission (as the portion off the stellar limb has a much smaller projected area). These configurations are understood to correspond, respectively, to what  was described as the `high' and `low' states of H$\alpha$ emission in other magnetic OB stars 
(e.g. \citealt{wade:2015}). In terms of putative longitudinal magnetic field variations, assuming a centred dipole, we only expect to see one magnetic pole as a result of our inferred $i$ and $\beta$ (a double-wave variation in H$\alpha$ would have been possible using a rotation period twice as long, as previously reported, leading to two maxima in emission per rotational cycle, as both magnetic pole would be seen -- however, we rule out this scenario, as discussed in Sect.\,\ref{section:rotation}). Therefore, at high state, the one magnetic pole that we do 
see is at its closest alignment to the line of sight; this is what we consider as phase 0, and it would yield a maximum value for $\langle B_\textrm{z} \rangle$. Conversely, at low state ($\phi = 0.5$), that same pole, while still visible, is now very near the limb of the star (and the other pole remains just out of sight), yielding a minimum value of $\langle B_\textrm{z} \rangle$.

That said, since $\xi$\,Per also exhibits more `classic' DAC behaviour (including cyclic variations in its \ion{Si}{IV} doublet, as shown in Fig.\,\ref{fig:xiSi4N4_87-94}), we also consider the presence of CIRs in the wind. While these \textit{normal} DACs are not typically found in the UV line profiles of strongly magnetic O-type stars, the weak confinement regime which we suspect applies to the wind of $\xi$\,Per might allow for the coexistence of a magnetosphere and CIRs. Specifically, CIR footpoints situated at moderate-to-high magnetic latitudes might lead to large-scale spiral structures that are qualitatively identical to those that are theorised to arise in the winds of non-magnetic stars, since as noted previously, for $\eta_* \lesssim 0.1$, the overall wind structure is largely unaffected beyond the region immediately surrounding the magnetic equator. Of course, this hypothesis would need to be confirmed by MHD simulations, but as this lies well beyond the scope of this current paper, we will simply assume it to be valid for the time being.

The picture emerging from this scenario is presented in Fig.\,\ref{fig:schematic}, 
showing the coexistence of both a disc of magnetospheric material and a CIR driven by a surface brightness spot. The spot is assumed to be sufficiently far from the magnetic equator so that the resulting CIR is largely unaffected by magnetic confinement. However, its longitude (and as a result, the rotational phase at which it appears) is a free parameter that can conceivably vary over timescales of a few to several rotation periods (as reported in e.g. $\zeta$\,Pup; \citealt{ramiaramanantsoa:2018}). Because the inner wind is largely dominated by the effect of the magnetic field, diagnostic lines that probe that region but are not sensitive to the outer wind (such as H$\alpha$ and the \ion{N}{iv} line, as shown in Fig.\,\ref{fig:xiSi4N4_87-94}) exhibit periodic variability, though some overlying stochastic variations can also appear as a result of CIRs (especially for H$\alpha$). On the other hand, diagnostics of the outer wind (such as the \ion{Si}{iv} doublet) exhibit variability that cannot be phased quite as clearly, and as a result, the signal appears to be smeared.

\begin{figure}[!t]
\begin{center}
\includegraphics[width=0.7\linewidth]{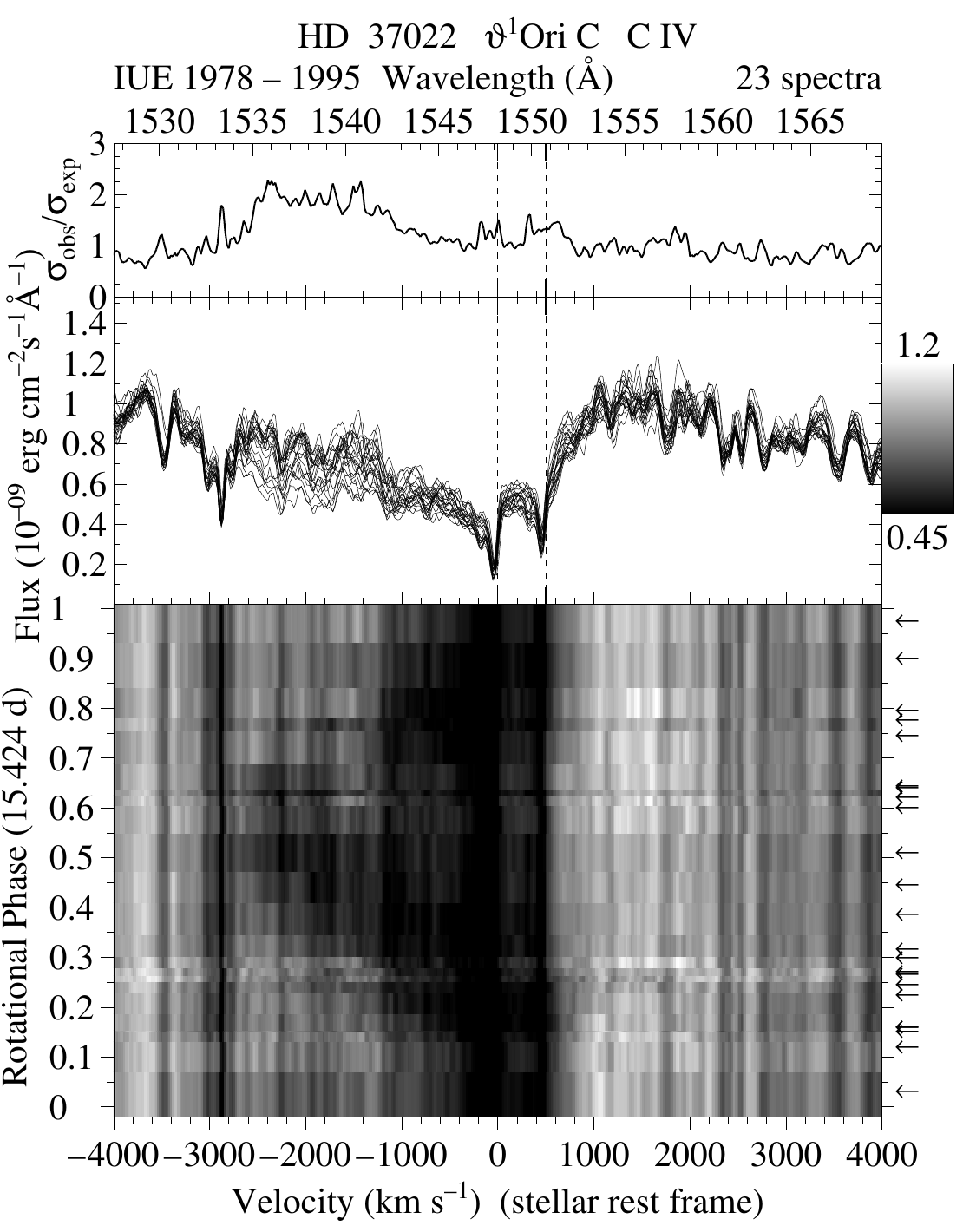}
\caption{Periodic variability in the \ion{C}{IV} doublet wind lines in the magnetic 
O7\,Vp star $\theta^1$\,Ori\,C observed during more than 16 years with 
IUE. \textit{Top}: Significance of variability, expressed as the ratio of the observed to the expected standard variation. Grey-scale plot shows the progression of the absorption as a function of phase in the 15.424\,d rotation period,
as collected arbitrarily sampled over 390 rotations. 
Phase 0 corresponds to maximum field strength at HJD 2448833.0.  
Note: the 
progression to high velocities is similar to that of the well-known DACs in presumably non-magnetic stars.
}
\label{fig:theta1OriCc4}
\end{center}
\end{figure}

This proposed phenomenology is also compatible with the behaviour of UV wind lines in 
$\theta^1$\,Ori\,C, the only magnetic O-type star to have phase-resolved UV spectroscopy obtained with IUE. 
However, in that case, since the magnetic field is significantly stronger (and consequently the star has a much larger $R_\textrm{A} = 2.4\,R_*$, \citealt{petit:2013}), it also affects lines that probe the outer wind.
In Fig.\,\ref{fig:theta1OriCc4}, in the same format as 
Fig.\,\ref{fig:SiNHedynamic}, we overplotted and 
phased-folded  
the \ion{C}{iv} $\lambda\lambda 1548,50$\,\AA\ doublet of $\theta^1$\,Ori\,C using 23 available IUE spectra with the rotational period of 15.242\,d. The dynamic plot is produced by dividing the spectra by a `least-absorption' template 
\citep{kaper:1997}. 
The wind periodicity is strictly conserved over 16 years, with two other significant issues: (1) the variability occurs mainly at high velocity, and
perhaps even more important, (2) the absorption features are 
more reminiscent of the classic DAC
structures 
observed in other (presumably non-magnetic) OB stars. For this star, the angles $i$ and $\beta$ are
both about 45$^{\circ}$ \citep{donati:2002}, which means that the north magnetic pole
points only once per rotation cycle towards the observer (at phase 0) whereas
at phase 0.5 the observer looks through the magnetic equator, a geometry that is quite similar to what we infer for $\xi$\,Per.

Finally, as mentioned earlier, strongly magnetised OB stars with large and dense magnetospheres (as epitomised by \object{NGC 1624-2}, \citealt{David-Uraz:2019a}) show dramatic changes in their P~Cygni line profiles as a function of rotational phase, with variations occurring both in emission and absorption. However, in this regard, the variation of $\xi$\,Per in its \ion{N}{IV} line is more reminiscent of the variations in the \ion{C}{iv} and \ion{Si}{iv} wind lines of later B-type magnetic stars, such as \object{$\beta$\,Cep} \citep{Henrichs:2013}, \object{$\zeta$\,Cas} \citep{neiner:2003a}, \object{V2052\,Oph} \citep{neiner:2003b}, and \object{$\sigma$\,Lup} \citep{henrichs:2012}. Their weaker winds lead to sparser magnetospheres, possibly making for a more adequate comparison with the small magnetosphere expected from very weak confinement around $\xi$\,Per. 

Regarding the spot(s) driving the `classic' DAC(s), we assume that results from previous studies that did not consider large-scale magnetic fields apply to the case of $\xi$\,Per as well. For instance,
the numerical simulations by \cite{david-uraz:2017} showed that only bright spots which cover a significant part of the surface can correctly reproduce the observed evolution of 
DACs. In their calculations, the spots were located in the rotational equatorial plane. For our adopted 
geometry, as illustrated in Fig.\,\ref{fig:schematic}, the spots will be located at higher latitudes with respect to the rotational axis, i.e.\ comprising a configuration which leads to the formation of DACs which is different from the equatorial case. In these cases, 3D models are needed to describe the time-dependent behaviour of the flow. This was done by \cite{dessart:2004} who studied the occurrence of CIRs in 3D models and gave specific examples of CIR formation which originates outside the equatorial plane, seemingly giving rise to large-scale spiral structures lying on a cone whose opening angle corresponds to the latitude of the spot from which the CIR was driven. (In this respect, we note that \cite{massa:2014} erroneously state that CIRs are thought to be confined to the equatorial plane.) Unfortunately, these non-equatorial CIR simulations were not coupled with radiative transfer calculations, so it is not yet known whether the DAC behaviour would differ from what was  inferred from simulations of equatorial CIRs. A 3D study of DACs by \cite{lobel:2008} for the B supergiant \object{HD\,64760}, which did include radiative transfer calculations, was restricted to the equatorial case.

Another observational constraint for modelling the UV variations 
in $\xi$\,Per is the phase bowing observed in the \ion{N}{IV} line 
\citep{kaper:1999}: the (negative) radial velocity at which the absorption shows up first, occurs at a higher velocity than the velocity at which the column density of the absorption reaches it maximum value. While for `classic' DACs, this is understood to stem from the fact that the structures in the wind are curved, and explains why material at a given velocity rotates into the line of sight first, followed by material moving at higher and lower velocities (
\citealt{owocki:1995}, \citealt{cranmer:1996}, and \citealt{dejong:2001}), the interpretation of this feature in the context of a large-scale magnetic field would require more careful modelling. Interestingly, all nine O stars with DACs show a decrease in phase towards higher velocities; only $\xi$\,Per also shows a lag in phase towards lower velocities.

\subsection{Investigating the origin of optical light variations}
\label{section:opticallight}

The origin of the periodic optical light variations is not immediately clear. It could be due to a similar obscuration effect as mentioned by \cite{massa:2014}, discussed 
earlier, to explain the covariability between H$\alpha$ and the X-ray light curve.

Our proposed magnetic geometry implies a single-wave modulation of the integrated wind column density as a function of rotational phase -- however, modelling the resulting optical light curve variations is beyond the scope of this paper.  
While this periodic broadband optical variability is expected in magnetic early-type stars (e.g. \citealt{David-uraz:2019b, Munoz:2020}), it can have very small amplitude in O-type stars and be dominated by other stochastic phenomena \citep{barron:2023}, making it more difficult to interpret, including its phasing. 
In particular, stochastic low-frequency (SLF) variability, which appears to be largely ubiquitous in massive stars, and is believed to be caused by internal gravity waves 
(e.g.\ \citealt{Bowman:2019}) or 
phenomena arising in the sub-surface convection zone (e.g.\  \citealt{Lecoanet:2019}). SLF variability results in an excess of power at low frequency, and in seemingly aperiodic variations evolving on characteristic timescales, not unlike what we see in Fig.\,\ref{fig:TESSLS}. The TESS sector 59 data of $\xi$\,Per  were analysed by \citet{Shen:2024}, who fit the SLF variability, finding a characteristic frequency of 3.13\,d$^{-1}$. 
Other sources of optical light variability include the above mentioned 
non-radial pulsations \citep{dejong:1999}, convection \citep{aerts:2015}, and wind blanketing \citep{krticka:2016, krticka:2018}. The latter effect occurs because of heating of the photosphere by wind-absorbed stellar flux, which means that a variable wind implies photometric variability. \cite{krticka:2018} estimated that for O-stars with stellar parameters similar to those of $\xi$\,Per, photometric changes of several millimagnitude 
amplitudes can indeed occur on hourly timescales, resembling real light curves of a number of O stars observed by the MOST and BRITE satellites. 
\cite{krticka:2016} predicts
that the UV flux variations at 1300\,\AA{} should behave in antiphase to the optical variations, 
which has still to be tested.  
In summary, at this stage it is clear that the photometric variability can be tied to the rotation period and hence could constrain the actual magnetic configuration. 

Finally, it is also possible that the dominant cyclic component of the variability contributes to the phase-averaged curve presented in Fig.\,\ref{fig:BRIGHTfit}. 
Photometric variations due to evolving 
photospheric drivers of large-scale wind structures (CIRs) and/or small-scale wind structures (clumps), as proposed for $\zeta$\,Pup \citep{ramiaramanantsoa:2018}, could be modelled using the parameters provided by the present paper. While these would likely cancel out over large timescales, that might not entirely be the case if 
these drivers occur at preferential longitudes or phases. Such a behaviour would be reminiscent of the so-called active longitudes that are observed on the Sun, which are longitudes where greater sunspot activity is observed, and are believed to be related to the large-scale component of the solar magnetic field (e.g. \citealt{berdyugina:2003}).
While there is no clear example of this behaviour in massive stars to this day, evolving light curve patterns over several rotational cycles in BRITE observations of $\zeta$\,Pup appear to suggest the possibility of spot distributions changing over time, but with photometric peaks clustering around specific phases \citep{ramiaramanantsoa:2018}. This could also explain the shift between the phase of maximum BRITE optical flux and the phase of maximum \ion{N}{IV} emission.

\subsection{Stochastic versus dipolar magnetic fields}
\label{stochastic}
\cite{cantiello:2009} showed that due to the 
recombination of iron in the outer layers of massive stars a subsurface convection zone arises inside the radiative envelope, which is theoretically believed to possibly give rise to dynamo action. This could lead to the generation of stochastic magnetic fields 
that rise to the surface via buoyancy on relatively short timescales, leading to bright magnetic spots 
\citep{braithwaite:2017}. Such spots are proposed to lead to CIRs \citep{cantiello:2011b}, and are compatible with the stochastic component of variability described above. Supporting evidence for the amplification of these buoyant magnetic fields is provided by MHD simulations by \cite{jiang:2017}, whereas stability considerations  were presented by \cite{sen:2018}. Suggested observational consequences include photometric variability 
(Sect.\,\ref{Section:photometry}), magnetic spots \citep{cantiello:2011b}, stellar prominences \citep{sudnik:2016}, excess X-ray emission \citep{waldron:2009, naze:2014, owocki:2016} and wind clumping \citep{puls:2006, sundqvist:2018}.
\cite{cantiello:2009} provide estimates of the turnover time and the strength of the short-lived 
fields generated by this process.  

A natural question is whether such stochastic fields are also present in stars with a dipole field, presumably with a fossil origin, as organised magnetic fields are inhibiting convection. \cite{braithwaite:2006a} showed that in a stable configuration, the poloidal field outside the star must be balanced by an internal toroidal field. When the magnetic energy of the stochastic component dominates over the toroidal component, the poloidal component cannot survive. Table~1 in \cite{cantiello:2009} provides the physical parameters to estimate the typical magnetic energy in the subsurface convection layer: $\rho v^2 \simeq 8.5 \times 10^4$\,erg. Setting this value to $ B^2/8\pi$ gives $B \approx 1.5 \times 10^3$\,G for a 20\,$M_{\odot}$ star with radius 10.46\,$R_{\odot}$, not so different from $\xi$\,Per, and which we will use as an example. (For a 60\,$M_{\odot}$ star with radius 22.04\,$R_{\odot}$ the field would be $\approx 2.7 \times 10^3$\,G.)  \cite{braithwaite:2009} gives an estimate of about 10 for the ratio of the toroidal to the surface polar field strength. This implies that a polar field stronger than 150\,G would be in competition with the generation of sub-surface fields. Such a star would have an about 3 times smaller observable value of $\langle B_{Z}\rangle$, depending on the values of $\beta$ and $i$.
These considerations are of course quantitatively uncertain, but may contain a source of further study on whether and under which conditions organised and stochastic fields could coexist, see for instance \cite{Sundqvist:2013}.

Later, the aforementioned competition  was  
suggested to lead to a `bistability' mechanism 
\citep{jermyn:2020}. In this scenario, large-scale fossil fields that are not strong enough to inhibit subsurface convection end up being destroyed by this convection, and eventually give rise to weak, dynamo-generated small-scale fields. Various additional theoretical prescriptions for the threshold strength that a fossil field needs to stabilise the atmosphere 
 were  formulated (e.g. \citealt{Sundqvist:2013,macdonald:2019}), but that threshold is generally consistent with the weakest large-scale fields found in massive and intermediate-mass stars \citep{auriere:2007}.

Assuming that the periodicity found in the \ion{N}{IV} line is indeed due to a stable, large-scale magnetic field on the surface of $\xi$\,Per, it could signify that the star is located in this `magnetic desert', making it an interesting test case for various hypotheses regarding the evolution and origin of magnetic fields in stars with radiative envelopes. If $\xi$\,Per indeed possesses a relatively weak large-scale field, it could therefore be in a transitional phase of its magnetic evolution, similarly perhaps to other fairly weakly magnetised evolved O-type stars (but to an even greater degree) like $\zeta$\,Ori\,Aa \citep{blazere:2015, buysschaert:2017} and 63\,Oph \citep{barron:2025}. \citet{jermyn:2020} mention that the timescale for the destruction of a fossil field below the critical strength can range between $\sim$50 and $10^4$ years, with more massive stars (in other words those with significant winds) being closer to the lower limit of that range. This would make $\xi$\,Per a potentially rare object. Such a weak field, coupled with $\xi$\,Per's luminosity class and likely evolved stage, also appears consistent with early indications from the LIFE survey that the distribution of magnetic field strengths in evolved stars peaks at a lower value than the same distribution in main-sequence stars (which is expected from simple flux conservation arguments  because these stars expand; e.g. \citealt{martin:2018}). 

Finally, given its status as a runaway star, it is also likely that $\xi$\,Per has undergone binary interaction at some point in its past (although alternatively it  was claimed that $\xi$\,Per became a runaway as a result of ejection during the initial relaxation of the \object{Perseus OB2 cluster}; \citealt{bhat:2022}). The topic of stellar mergers has gained a lot of attention in recent years as a possible origin for fossil magnetic fields in massive stars (e.g. \citealt{Schneider:2016}). Recent observational studies have corroborated the idea that binary interactions more generally might lead to the generation of a magnetic field, not only for mergers (as suspected in the case of \object{HD 148937}, \citealt{frost:2024}; and for the extremely strongly magnetised helium star \object{HD\,45166}, \citealt{shenar:2023}), but also possibly in cases of mass transfer (as is suspected for the magnetic component in HD\,47129, also known as Plaskett's star; Wade et al. 2025, priv.\ comm.). If $\xi$\,Per were to be detected as hosting a fossil-type magnetic field, it would likely be a strong candidate for the latter case, by virtue of its relatively rapid rotation and runaway status.

\subsection{Period search in other O stars}
We searched for periodicity with similar techniques as described above in the next three best studied O stars with high-resolution spectra in the IUE archive, which stretches 
more than 15 years. These are \object{68 Cyg} O7.5III:n((f)), \object{$\zeta$\,Pup} O4If, and \object{$\lambda$ Cep} O6I(n)fp with 228, 219, and 179 spectra, respectively (noting that for $\xi$\,Per, this number is 326). No significant periodicity could be detected near their estimated rotation periods. This is, however, not compelling evidence that the studied wind lines are not probing the rotation, since the number of spectra, the S/N, or both are smaller than for $\xi$\,Per, which would make a detection of such a faint signal among very unequally spaced data points much more challenging. So the question whether other O stars without a detected magnetic field may still host a corotating dipolar field can therefore not be answered at this stage.

\section{Conclusions}
\label{section:conclusions}
We have shown that over the time span of
33 years, the low-velocity absorption part of the \ion{N}{IV} $\lambda$1718\,\AA\ line of $\xi$\,Per has been shown to contain a prominent  sinusoidal signal with a period of 2.040514 $\pm$ 0.000020\,d, which we identify as the stellar rotation period.
This result stands in opposition to earlier conclusions favouring a rotation period of $\sim$4\,d, which was based on the assumption of a much larger stellar radius than we adopted in this work, following from a recent interferometric measurement of the angular diameter of the star. All earlier reports of blue-wing absorption variations in the H$\alpha$ line are consistent with this period and have the same phase dependency, which confirms that these lines were formed in the same region, just above the stellar surface. The most likely origin of the rotational modulation is the presence of a weak (presumably) dipolar magnetic field, which has nonetheless remained undetected thus far. We note that all previous spectropolarimetric measurements  were made at limited, unfavourable, rotational phases. We provide the ephemeris at which new measurements will be able to better constrain the amplitude of a possible magnetic field. 

Crucially, we also distinguish between the strict periodicity seen in the \ion{N}{IV} low-velocity flux variations and the apparently cyclic nature of the DACs found in the \ion{Si}{IV} doublet, especially at high velocities. We still consider the latter to be consistent with the prevalent CIR model and propose that the spots that drive these CIRs originate from moderate-to-high magnetic latitudes; therefore, we find that the resulting CIRs are largely unaffected by the putative large-scale magnetic field. This leads us to propose a toy model that involves both a magnetospheric disc with a varying projected area and the aforementioned CIRs, all contributing to the variability seen across multiple observational diagnostics, probing both the inner and outer wind, as illustrated in Fig.\,\ref{fig:schematic}.
Such a model could qualitatively satisfy a number of observational constraints, in particular, the 
dependency of the variations across the \ion{N}{IV}, \ion{Si}{IV}, \ion{He}{II}, and H$\alpha$ lines.
While the X-ray flux measurements do seem to phase in accordance with our toy model, it is unclear whether the optical photometry (especially from the BRITE mission) follows a similar phasing, as it might be dominated by stochastic co-rotating surface bright spots;  however,  we note that the former is still insufficient to constrain such a model,  We discuss the inferred magnetic properties of $\xi$\,Per in the light of 
the existing observational upper limit \citep{david-uraz:2014} and conclude that any large-scale magnetic field 
below that limit would only produce weak wind confinement. We also discuss the potential interplay between such a large-scale field and the possible generation of stochastic, small-scale fields, the likes of which  were proposed as the physical drivers for CIRs. 
We could not detect a periodicity in IUE spectra of three other bright O stars, possibly because their datasets are severely limited in quality, number, and/or coverage; alternatively it could be because $\xi$\,Per's magnetic status is unique. Highly sensitive magnetic measurements in the future might eventually prove or disprove our conjecture that $\xi$\,Per hosts a magnetic field, while shedding light on how the magnetic surface properties interact with the photospheric drivers of the DACs.

{
\section*{Data availability}
\label{electronic}
Tables \ref{tab:IUEtable} and \ref{tab:STIStable} are only available in electronic form at the CDS via anonymous ftp to cdsarc.u-strasbg.fr (130.79.128.5) or via \url{http://cdsweb.u-strasbg.fr/cgi-bin/qcat?J/A+A/ .}
}

\begin{acknowledgements}
We are very grateful to the observers who contributed to this project, in particular to Douglas Gies at Kitt Peak and Regina Zwarthoed at Calar Alto during the 1989 campaign. Special thanks go to Katie Gordon and Douglas Gies for providing the angular diameter of $\xi$\,Per, and to Tahina Ramiaramanantsoa for 
helpful discussions and comments regarding results of MOST and BRITE photometry. 
We gratefully acknowledge insightful discussions with Lex Kaper, Henk Spruit, Asif ud-Doula,  Alex de Koter, Zsolt Keszthelyi,  
Sarah Brands, and David Cohen. We are also much indebted to Stan Owocki 
for helpful correspondence and reading a first version of the manuscript. Jari van Opijnen kindly provided the corrected Gaia distance value.  We gratefully thank the late
Derck Massa for his generous comments on the analysis of STIS spectra. We thank Joy Nichols and Myron Smith for commenting on the wavelength shift issue between the INES and MAST archival spectra, and Ralph Bohlin for sharing his expertise on absolute UV flux calibrations. 
We thank Alex Fullerton for carefully sorting out the issue clarified in Fig.\,\ref{fig:Chandra} and for discussions. We gratefully acknowledge the clarifying comments by Daniela Huppenkothen on confidence levels in Lomb-Scargle periodograms. 
NS fondly thanks the warm hospitality and partial financial support of the Anton Pannekoek Institute for Astronomy, where most of this work was done. She is also grateful to NOVA, the Leids Kerkhoven Bosscha Fonds, and to Lex Kaper, who provided additional financial support. ADU also gratefully acknowledges financial support from Lex Kaper, as well as support from the \textit{Chandra} General Observer Program via award TM2-23001A and from NASA
ROSES award 22-ADAP22-0070.
HFH warmly acknowledges Lex Kaper for his continuous support, also financially, during the course of this extensive project, which  started in 2017. HFH expresses special thanks to Stan Owocki for his suggestion to invite ADU as a coauthor without whom this paper would have never been completed. The authors are grateful to an anonymous referee for suggesting improvements, in particular regarding the run-away status of $\xi$\,Per and its distance determination.  This research has made use of the SIMBAD database, operated at CDS, Strasbourg, France. 
This work is based {in part} on observations made with the NASA/ESA \textit{Hubble} Space Telescope, and obtained from the Hubble Legacy Archive. 
The data were obtained from the Mikulski Archive for Space Telescopes at the Space Telescope Science Institute, which is operated by the Association of Universities for Research in Astronomy, Inc., under NASA contract NAS 26555 for HST. These observations are associated with program 14180 (PI D. Massa). 
This paper includes data collected with the TESS mission, obtained from the MAST data archive at the Space Telescope Science Institute (STScI). Funding for the TESS mission is provided by the NASA Explorer Program. 
Parts of Sect.\,\ref{Section:photometry} are based on data collected by the BRITE Constellation satellite mission, designed, built, launched, operated and supported by the Austrian Research Promotion Agency (FFG), the University of Vienna, the Technical University of Graz, the Canadian Space Agency (CSA), the University of Toronto Institute for Aerospace Studies (UTIAS), the Foundation for Polish Science $\&$ Technology (FNiTP MNiSW), and the National Science Centre (NCN).  Figs.\,\ref{fig:xiSi4N4_87-94}, \ref{fig:SiNHedynamic}, and \ref{fig:periodogram} were programmed in Adobe PostScript\textregistered\      level 2. 
This work made use of {Astropy\footnote{\url{http://www.astropy.org}}}: a community-developed core Python package and an ecosystem of tools and resources for astronomy \citep{astropy:2022}.
\end{acknowledgements}

\bibliographystyle{aa}
\bibliography{references}

\end{document}